\documentclass[format=acmsmall, review=false, screen=true]{acmart}

\usepackage{booktabs} 

\usepackage{subcaption,caption}
\usepackage{ragged2e}
\usepackage{array}
\usepackage{url}
\usepackage{multirow}
\usepackage{colortbl}
\usepackage{lipsum}

\newcolumntype{P}[1]{>{\centering\arraybackslash}p{#1}}

\usepackage{color}
\usepackage{soul}
\definecolor{grannysmithapple}{rgb}{0.66, 0.89, 0.63}

\usepackage{lscape}
\usepackage{graphicx}
\usepackage{wrapfig}
\usepackage{rotating}
\usepackage{epstopdf}

\begin{document}
\title[A Survey on Recent Hardware Data Prefetching Approaches with An Emphasis on Servers]{A Survey on Recent Hardware Data Prefetching Approaches with An Emphasis on Servers}  
\author{Mohammad~Bakhshalipour}
\orcid{0000-0001-6648-1773}
\affiliation{%
  \institution{Sharif University of Technology}
  \department{Department of Electrical Engineering}
  \streetaddress{Azadi}
  \city{Tehran}
  \state{Tehran}
  \postcode{11155-9517}
  \country{Iran}}
\author{Mehran~Shakerinava}
\affiliation{%
  \institution{Sharif University of Technology}
  \department{Department of Electrical Engineering}
  \streetaddress{Azadi}
  \city{Tehran}
  \state{Tehran}
  \postcode{11155-9517}
  \country{Iran}}
\author{Fatemeh~Golshan}
\affiliation{%
  \institution{Sharif University of Technology}
  \department{Department of Electrical Engineering}
  \streetaddress{Azadi}
  \city{Tehran}
  \state{Tehran}
  \postcode{11155-9517}
  \country{Iran}}
\author{Ali~Ansari}
\affiliation{%
  \institution{Sharif University of Technology}
  \department{Department of Electrical Engineering}
  \streetaddress{Azadi}
  \city{Tehran}
  \state{Tehran}
  \postcode{11155-9517}
  \country{Iran}}
\author{Pejman~Lotfi-Kamran}
\affiliation{%
  \institution{Institute for Research in Fundamental Sciences (IPM)}
  \department{School of Computer Science}
  \city{Tehran}
  \state{Tehran}
  \postcode{19538-33511}
  \country{Iran}}
\author{Hamid~Sarbazi-Azad}
\affiliation{%
  \institution{Sharif University of Technology \& Institute for Research in Fundamental Sciences (IPM)}
  \department{Department of Computer Engineering \& School of Computer Science}
  \city{Tehran}
  \state{Tehran}
  \postcode{19538-33511}
  \country{Iran}
}

\newcommand\kasraa[1]{\noindent{\color{red} {\bf \fbox{Kasraa}} {\it#1}}}
\newcommand\TODO[1]{\noindent{\color{blue} {\bf \fbox{TODO}} {\it#1}}}
\newcommand\methodname[1]{\textsc{{#1}}}
\newcommand\appname[1]{\textsf{{#1}}}
\newcommand\componentname[1]{\textsl{{#1}}}
\newcommand{\pluseq}{\mathrel{+}=}

\begin{abstract}
Data prefetching, i.e., the act of predicting application's future memory accesses and fetching those that are not in the on-chip caches, is a well-known and widely-used approach to hide the long latency of memory accesses. The fruitfulness of data prefetching is evident to both industry and academy: nowadays, almost every high-performance processor incorporates a few data prefetchers for capturing various access patterns of applications; besides, there is a myriad of proposals for data prefetching in the research literature, where each proposal enhances the efficiency of prefetching in a specific way.  In this survey, we discuss the fundamental concepts in data prefetching and study state-of-the-art hardware data prefetching approaches.

\end{abstract}

%
%
%
%



\thanks{ 

  Author's addresses: Mohammad~Bakhshalipour, Department of Computer Engineering, Sharif University of Technology;
  Mehran~Shakerinava, Department of Computer Engineering, Sharif University of Technology;
  Fatemeh~Golshan, Department of Computer Engineering, Sharif University of Technology;
  Ali~Ansari, Department of Computer Engineering, Sharif University of Technology;
  Pejman~Lotfi-Kamran, School of Computer Science, Institute for Research in Fundamental Sciences (IPM); 
  Hamid~Sarbazi-Azad, Department of Computer Engineering, Sharif University of Technology and School of Computer Science, Institute for Research in Fundamental Sciences (IPM)
}

\maketitle

\renewcommand{\shortauthors}{M.~Bakhshalipour et al.}

\section{Introduction}
\label{chapter:introduction}

Server workloads like \componentname{Media Streaming} and \componentname{Web Search} serve millions of users and are considered an important class of applications. Such workloads run on large-scale data-center infrastructures that are backed by processors which are essentially tuned for low latency and quality-of-service guarantees. These processors typically include a handful of high-clock frequency, aggressively-speculative, and deeply-pipelined cores so as to run server applications as fast as possible, satisfying end-users' latency requirements~\cite{Lotfi-Kamran:2012:SP:2337159.2337217, Grot:2012:ODT:2412372.2412724, Lim:2008:UDN:1381306.1382148, Ferdman:2012:CCS:2150976.2150982, esmaili2018scale, ansari2019code, mireshghallah2019energy, jokar2020baldur, jokar2019high, al2008scalable, farrington2010helios}. 

Much to processor designer's chagrin, bottlenecks in the memory system prevent server processors from getting high performance on server applications. As server workloads operate on a large volume of data, they produce active memory working sets that dwarf the capacity-limited on-chip caches of server processors and reside in the off-chip memory; hence, these applications frequently miss the data in the on-chip caches and access the long-latency memory to retrieve it. Such frequent data misses preclude server processors from reaching their peak performance because cores are idle waiting for the data to arrive~\cite{Ferdman:2012:CCS:2150976.2150982, Lotfi-Kamran:2012:SP:2337159.2337217, Karkhanis:2004:FSP:1028176.1006729, Hameed:2010:USI:1815961.1815968, Ferdman:2012:QME:2382553.2382557, vakil2018cache, ghahani2018making, Bakhshalipour:2019:EHD:3341324.3312740, bakhshalipour2019reducing, livia, esfeden2019corf, khorasani2018register, khorasani2018regmutex, Kayaalp_RRI, hojabr2017customizing}. 

System architects have proposed various strategies to overcome the performance penalty of frequent memory accesses. \componentname{Data Prefetching} is one of these strategies that has demonstrated significant performance potentials. Data prefetching is the art of predicting future memory accesses and fetching those that are not in the cache before a core explicitly asks for them in order to \emph{hide} the long latency of memory accesses. Nowadays, virtually every high-performance computing chip uses a few data prefetchers (e.g., \componentname{Intel Xeon Phi}~\cite{knights_landing}, \componentname{IBM Blue Gene/Q}~\cite{haring2012ibm}, \componentname{AMD Opteron}~\cite{amd_opteron}, and \componentname{UltraSPARC III}~\cite{horel1999ultrasparc}) to capture regular and/or irregular memory access patterns of various applications. In the research literature, likewise, there is a myriad of proposals for data prefetching, where every proposal makes the prefetching more efficient in a specific way. 

In this study, we first discuss the fundamental concepts in data prefetching then study recent, as well as classic, hardware data prefetchers in the context of server workloads. We describe the operations of every data prefetcher, in detail, and shed light on its design trade-offs. In a nutshell, we make the following contributions in this study:

\begin{itemize}

\item We describe memory access patterns of applications and discuss how these patterns lead to different classes of correlations, from which data prefetchers can predict future memory accesses. 

\item We describe the operations of state-of-the-art hardware data prefetchers in the research literature and discuss how they are able to capture data cache misses. 

\item We highlight the overheads of every data prefetching technique and discuss the feasibility of implementing it in modern processors.

\end{itemize}

\subsection{Why Hardware Data Prefetching?}
\label{chapter:introduction:other_than_hardware_prefetching_methods}

Progress in technology fabrication accompanied by circuit-level and microarchitectural advancements have brought about significant enhancements in the processors' performance over the past decades. Meanwhile, the performance of memory systems has not held speed with that of the processors, forming a large gap between the performance of processors and memory systems~\cite{Wulf:1995:HMW:216585.216588, Trancoso:1997:MPD:548716.822671, Ailamaki:1999:DMP:645925.671662, Hankins:2003:SCR:956417.956541, hardavellas:database, bakhshalipour2018stacked, Rashidi:2018:IMP:3199680.3177965, saeed_csur, nesta_mirzaeian, tcd_mirzaeian, armin_dsm, jeon2019locality, jokar2018cooperative}. As a consequence, numerous approaches have been proposed to enhance the execution performance of applications by bridging the processor-memory performance gap. Hardware data prefetching is just one of these approaches. Hardware data prefetching bridges the gap by proactively fetching the data ahead of the cores' requests in order to eliminate the idle cycles in which the processor is waiting for the response of the memory system. In this section, we briefly review the other approaches that target the same goal (i.e., bridging the processor-memory performance gap) but in other ways.

\textbf{Multithreading}~\cite{Nemirovsky:2013:MA:2502821} enables the processor to better utilize its computational resources, as stalls in one thread can be overlapped with the execution of other thread(s)~\cite{Ryoo_OPA, akkary1998dynamic, Cui:2010:SDM:1924943.1924958, bakhshalipour2018parallelizing}. Multithreading, however, only improves \emph{throughput} and does nothing for (or even worsens) the \emph{response time}~\cite{hardavellas:database, Lo:1998:ADW:279358.279367, Lotfi-Kamran:2012:SP:2337159.2337217}, which is crucial for satisfying the strict latency requirements of server applications.

\textbf{Thread-Based Prefetching} techniques~\cite{Collins:2001:SPL:379240.379248, Ganusov:2006:FEP:1187976.1187979, Lee:2009:PHT:1591896.1592265, Kamruzzaman:2011:IPM:1950365.1950411, Collins:2001:DSP:563998.564037} exploit idle thread contexts or distinct pre-execution hardware to drive helper threads that try to overlap the cache misses with speculative execution. Such helper threads, formed either by the hardware or by the compiler, execute a piece of code that prefetches for the main thread. Nonetheless, the \emph{additional} threads and fetch/execution bandwidth may not be available when the processor is fully utilized. The abundant request-level parallelism of server applications~\cite{Ferdman:2012:CCS:2150976.2150982, Lotfi-Kamran:2012:SP:2337159.2337217} makes such schemes ineffective in that the helper threads need to compete with the main threads for the hardware context. 

\textbf{Runahead Execution}~\cite{Mutlu:2003:REA:822080.822823, Mutlu:2005:TEP:1069807.1070000} makes the execution resources of a core that would otherwise be stalled on an off-chip cache miss to go ahead of the stalled execution in an attempt to discover additional load misses. 
Similarly, \textbf{Branch Prediction Directed Prefetching}~\cite{Kadjo:2014:BBP:2742155.2742218} utilizes the branch predictor to run in advance of the executing program, thereby prefetching load instructions along the expected future path. 
Such approaches, nevertheless, are \emph{constrained} by the accuracy of the branch predictor and can cover simply a portion of the miss latency, since the runahead thread/branch predictor may not be capable of executing far ahead in advance to \emph{completely} hide a cache miss. Moreover, these approaches can only prefetch \emph{independent} cache misses~\cite{Hashemi:2016:ADC:3001136.3001184} and may not be effective for many of the server workloads, e.g., \emph{OLTP} and \emph{Web} applications, that are characterized by long chains of dependent memory accesses~\cite{Ranganathan:1998:PDW:291069.291067, bakhshalipour2018domino}. 

On the software side, there are efforts to re-structure programs to boost chip-level \textbf{Data Sharing} and \textbf{Data Reuse}~\cite{Johnson:2007:SS:1325851.1325894, Larus:2002:UCE:647057.713864} in order to decrease off-chip accesses. While these techniques are useful for workloads with modest datasets, they fall short of efficiency for big-data server workloads, where the multi-gigabyte working sets of workloads dwarf the few megabytes of on-chip cache capacity. The ever-growing datasets of server workloads make such approaches \emph{unscalable}. \textbf{Software Prefetching} techniques~\cite{Zhang:2006:SPE:1121992.1122392, Luk:1996:CPR:237090.237190, Roth:1999:EJP:300979.300989, Chilimbi:2002:DHD:512529.512554,Chen:2007:IHJ:1272743.1272747} profile the program code and insert prefetch instructions to eliminate cache misses. While these techniques are shown to be beneficial for small benchmarks, they usually require significant \emph{programmer effort} to produce optimized code to generate timely prefetch requests.

\textbf{Memory-Side Prefetching} techniques~\cite{Hughes:2005:MPL:1066486.1066491, Solihin:2002:UUM:545215.545235, Yedlapalli:2013:MMI:2523721.2523761} place the hardware for data prefetching near DRAM, for the sake of saving precious SRAM budget. In such approaches (e.g.,~\cite{Solihin:2002:UUM:545215.545235}), prefetching is performed by a user thread running near the DRAM, and prefetched pieces of data are sent to the on-chip caches. Unfortunately, such techniques lose the \emph{predictability} of core requests~\cite{Mittal:2016:SRP:2966278.2907071} and are incapable of performing \emph{cache-level optimizations} (e.g., avoiding cache pollution~\cite{Srinath:2007:FDP:1317533.1318101}).

\subsection{Background}

\label{chapter:introduction:background}

In this section, we briefly overview a background on hardware data prefetching and refer the reader to prior work~\cite{Falsafi:2014:PHP:2643033, Mittal:2016:SRP:2966278.2907071, Bakhshalipour:2019:EHD:3341324.3312740} for more details. For simplicity, in the rest of the manuscript, we use the term \emph{prefetcher} to refer to the \textit{core-side hardware data prefetcher}.

\subsubsection{Predicting Memory References}

The first step in data prefetching is \emph{predicting} future memory accesses. Fortunately, data accesses demonstrate several types of correlations and localities, that lead to the formation of \emph{patterns} among memory accesses, from which data prefetchers can predict future memory references. These patterns emanate from the layout of programs' data structures in the memory, and the algorithm and the high-level programming constructs that operate on these data structures.

In this work, we classify the memory access patterns of applications into three distinct categories: (1) \emph{strided}, (2) \emph{temporal}, and (3) \emph{spatial} access patterns. \\


\noindent\textbf{Strided Accesses:}
\label{chapter:introduction:background:prediction:stride}
Strided access pattern refers to a sequence of memory accesses in which the \emph{distance} of consecutive accesses is constant (e.g., $\{A, A+k, A+2k, \ldots{}\}$). Such patterns are abundant in programs with dense matrices and frequently come into sight when programs operate on multi-dimensional arrays. Strided accesses also appear in pointer-based data structures when memory allocators arrange the objects sequentially and in a constant-size manner in the memory~\cite{Dahlgren:1995:EHS:527072.822612}.\\

\noindent\textbf{Temporal Address Correlation:}
\label{chapter:introduction:background:prediction:temporal}
Temporal address correlation~\cite{bakhshalipour2018domino} refers to a sequence of addresses that favor being accessed together and in the same \emph{order} (e.g., if we observe $\{A, B, C, D\}$, then it is likely for $\{B, C, D\}$ to follow $\{A\}$ in the future). Temporal address correlation stems fundamentally from the fact that programs consist of loops, and is observed when data structures such as lists, arrays, and linked lists are traversed. When data structures are stable~\cite{Chilimbi:2001:STD:645988.674166}, access patterns recur, and the temporal address correlation is manifested~\cite{bakhshalipour2018domino}.\\

\noindent\textbf{Spatial Address Correlation:}
\label{chapter:introduction:background:prediction:spatial}
Spatial address correlation~\cite{bakhshalipour2019bingo} refers to the phenomenon that similar access patterns occur in different \emph{regions} of memory (e.g., if a program visits locations $\{A, B, C, D\}$ of Page $X$, it is probable that it visits locations $\{A, B, C, D\}$ of other pages as well). Spatial correlation transpires because applications use various objects with a regular and fixed layout, and accesses reappear while traversing data structures~\cite{bakhshalipour2019bingo}.\\

\subsubsection{Prefetching Lookahead}
\label{chapter:introduction:background:lookahead}

Prefetchers need to issue \emph{timely} prefetch requests for the predicted addresses. Preferably, a prefetcher sends prefetch requests well in advance and supply enough storage for the prefetched blocks in order to hide the entire latency of memory accesses. An early prefetch request may cause evicting a useful block from the cache, and a late prefetch may decrease the effectiveness of prefetching in that a portion of the long latency of a memory access is exposed to the processor.

Prefetching lookahead refers to \emph{how far ahead of the demand miss stream the prefetcher can send requests}. An aggressive prefetcher may offer a high prefetching lookahead (say, eight) and issue many prefetch requests ahead of the processor to hide the entire latency of memory accesses; on the other hand, a conservative prefetcher may offer a low prefetching lookahead and send a single prefetch request in advance of the processor's demand to avoid wasting resources (e.g., cache storage and memory bandwidth).  Typically, there is a trade-off between the aggressiveness of a prefetching technique and its accuracy: making a prefetcher more aggressive usually leads to covering more data-miss--induced stall cycles but at the cost of fetching more useless data.

Some pieces of prior work propose to dynamically \emph{adjust} the prefetching lookahead~\cite{Srinath:2007:FDP:1317533.1318101, Kadjo:2014:BBP:2742155.2742218, Kim:2016:PCB:3195638.3195711}. Based on the observation that the optimal prefetching degree is different for various applications and various execution phases of a particular application, as well, these approaches employ heuristics to increase or decrease the prefetching lookahead, dynamically at run-time. For example, \methodname{SPP}~\cite{Kim:2016:PCB:3195638.3195711} monitors the accuracy of issued prefetch requests and reduce the prefetching lookahead if the accuracy becomes smaller than a predefined threshold. 

\subsubsection{Location of Data Prefetcher}
Prefetching can be employed to move the data from lower levels of the memory hierarchy to any higher level\footnote{We use the term higher (lower) levels of the memory hierarchy to refer to the levels closer to (further away from) the core, respectively.}. Prior work used data prefetchers at all cache levels, from the primary data cache to the shared last-level cache.

The location of a data prefetcher has a profound impact on its overall behavior~\cite{Mehta:2014:MCP:2597652.2597660}. A prefetcher in the first-level cache can observe  all memory accesses, and hence, is able to issue highly-accurate prefetch requests, but at the cost of imposing large storage overhead for recording the metadata information. In contrast, a prefetcher in the last-level cache observes the access sequences that have been filtered at higher levels of the memory hierarchy, resulting in lower prediction accuracy, but higher storage efficiency. 

\subsubsection{Prefetching Hazards}
\label{chapter:introduction:background:hazards}
A naive deployment of a data prefetcher not only may not improve the system performance but also may significantly harm the performance and energy-efficiency~\cite{bakhshalipour2018fast}. The two well-known major drawbacks of data prefetching are (1) cache pollution and (2) off-chip bandwidth overhead.\\

\noindent\textbf{Cache Pollution:}
Data prefetching may increase the demand misses by replacing useful cache blocks with useless prefetched data, harming the performance. Cache pollution usually occurs when an aggressive prefetcher exhibits low accuracy and/or when prefetch requests of a core in a many-core processor compete for shared resources with demand accesses of other cores~\cite{Ebrahimi:2009:CCM:1669112.1669154}. \\

\noindent\textbf{Bandwidth Overhead:}
In a many-core processor, prefetch requests of a core can delay demand requests of another core because of contending for memory
bandwidth~\cite{Ebrahimi:2009:CCM:1669112.1669154}. This interference is the major obstacle of using data prefetchers in many-core processors, and the problem gets thornier as the number of cores increases~\cite{Ebrahimi:2010:FVS:1736020.1736058,kim2010atlas}.\\


\subsubsection{Placing Prefetched Data}

Data prefetchers usually place the prefetched data into one of the following two structures: (1) the cache itself, and (2) an auxiliary buffer next to the cache. In case an auxiliary buffer is used for the prefetched data, demand requests first look for the data in the cache; if the data is not found, the auxiliary buffer is searched before sending a request to the lower levels of the memory hierarchy.  

Storing the prefetched data into the cache lowers the latency of accessing data when the prediction is correct. However, when the prediction is incorrect or when the prefetch request is not timely (i.e., too early), having the prefetched data in the cache may result in evicting useful cache blocks.\\

\subsection{A Preliminary Hardware Data Prefetcher}
\label{chapter:introduction:ibsp}

To give insight on how a stereotype operates, now we describe a preliminary-yet-prevalent type of stride prefetching. Generally, stride prefetchers are widely used in commercial processors (e.g., \componentname{IBM Power4}~\cite{tendler2002power4}, \componentname{Intel Core} ~\cite{doweck2006inside}, \componentname{AMD Opteron}~\cite{amd_opteron}, \componentname{Sun UltraSPARC III}~\cite{horel1999ultrasparc}) and have been shown quite effective for desktop and engineering applications. Stride prefetchers~\cite{Baer:1991:EOP:125826.125932, Sherwood:2000:PSB:360128.360135, Ishii:2009:AMP:1542275.1542349, Sair:2003:DPS:642791.642793, Jouppi:1990:IDC:325164.325162, Palacharla:1994:ESB:191995.192014, Zhang:2000:HSP:335231.335247, Iacobovici:2004:ESE:1006209.1006211} detect streams (i.e., the sequence of consecutive addresses) that exhibit strided access patterns (Section~\ref{chapter:introduction:background:prediction:stride}) and generate prefetch requests \emph{by adding the detected stride to the last observed address}. 

\begin{figure}[h]
   \centering
   \includegraphics[width=0.7\textwidth]{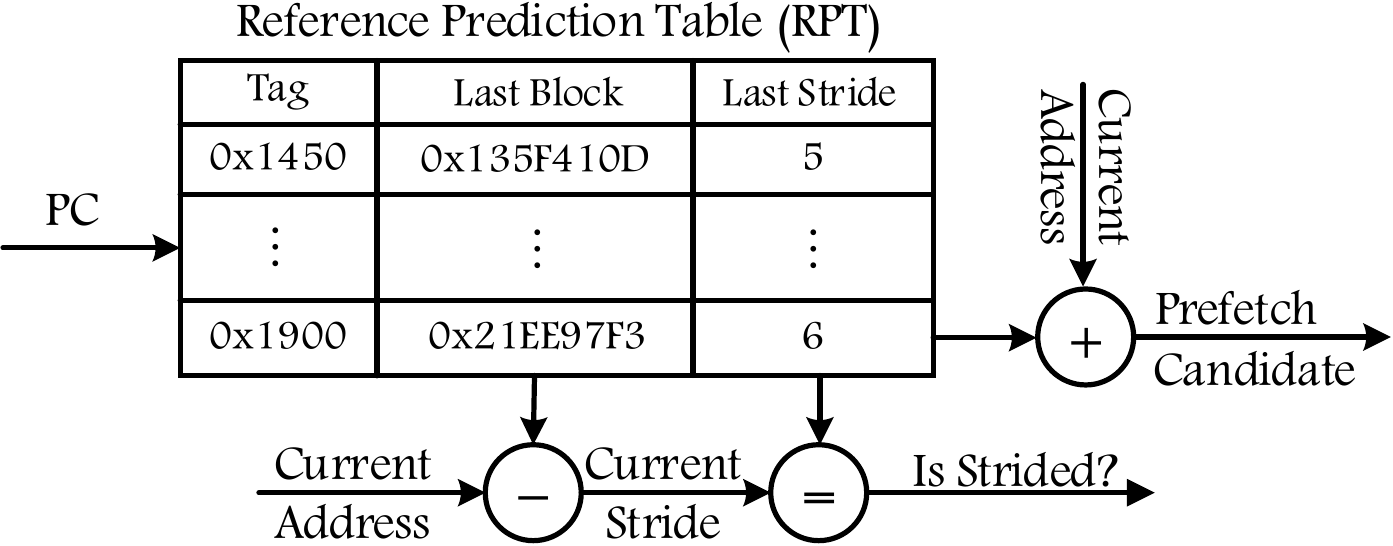}
   \caption{The organization of Instruction-Based Stride Prefetcher (IBSP). The `RPT' keeps track of various streams.
   \label{fig:stride-diagram}}
\end{figure}

\methodname{Instruction-Based Stride Prefetcher (IBSP)}~\cite{Baer:1991:EOP:125826.125932} is a preliminary type of stride prefetching. The prefetcher tracks the strided streams on a per load instruction basis: the prefetcher observes accesses issued by individual load instructions and sends prefetch requests if the accesses manifest a strided pattern. Figure~\ref{fig:stride-diagram} shows the organization of \methodname{IBSP}'s metadata table, named \componentname{Reference Prediction Table (RPT)}. \componentname{RPT} is a structure tagged and indexed with the \emph{Program Counter (PC)} of load instructions. Each entry in the \componentname{RPT} corresponds to a specific load instruction; it keeps the \componentname{Last Block} referenced by the instruction and the \componentname{Last Stride} observed in the stream (i.e., the distance of two last addresses accessed by the instruction).

Upon each trigger access (i.e., a cache miss or a prefetch hit), the \componentname{RPT} is searched with the PC of the instruction. If the search results in a miss, it means that no history does exist for the instruction, and hence, no prefetch request can be issued. Under two circumstances, a search may result in a miss: (1) whenever a load instruction is a new one in the execution flow of the program, and ergo, no history has been recorded for it so far, and (2) whenever a load instruction is re-executed after a long time, and the corresponding recorded metadata information has been evicted from the \componentname{RPT} due to conflicts. In such cases when no matching entry does exist in the \componentname{RPT}, a new entry is allocated for the instruction, and possibly a victim entry is evicted. The new entry is tagged with the PC, and the \componentname{Last Block} field of the entry is filled with the referenced address. The \componentname{Last Stride} is also set to zero (an invalid value) as no stride has yet been observed for this stream. However, if searching the \componentname{RPT} results in a hit, it means that there is a recorded history for the instruction. In this case, the recorded history information is checked with the current access to find out whether or not the stream is a strided one. To do so, the difference of the current address and the \componentname{Last Block} is calculated to get the \emph{current stride}. Then, the current stride is checked against the recorded \componentname{Last Stride}. If they do not match, it is implied that the stream does not exhibit a strided access pattern. However, if they match, it is construed that the stream is a strided one as three consecutive accesses have produced two \emph{identical} strides. In this case, based on the lookahead of the prefetcher (Section~\ref{chapter:introduction:background:lookahead}), several prefetch requests are issued by \emph{consecutively} adding the observed stride to the requested address. For example, if the current address and the current stride are $A$ and $k$, respectively, and the lookahead of prefetching is three, prefetch candidates will be $\{A+k, A+k+k, A+k+k+k\}$. Finally, regardless of the fact that the stream is strided or not, the corresponding \componentname{RPT} entry is updated: the \componentname{Last Block} is updated with the current address, and the \componentname{Last Stride} takes the value of the current stride. \\

In the following chapters, we introduce state-of-the-art data prefetchers and describe their mechanism.

\section{Spatial Prefetching}

Spatial data prefetchers predict future memory accesses by relying on spatial address correlation, i.e., the similarity of access patterns among multiple \emph{regions} of memory. Access patterns demonstrate spatial correlation because applications use data objects with a regular and fixed layout, and accesses reoccur when data structures are traversed~\cite{bakhshalipour2019bingo}. Spatial data prefetchers~\cite{bakhshalipour2019bingo, Somogyi:2006:SMS:1135775.1136508, Nesbit:2004:DCP:1072448.1072460, Shevgoor:2015:EPC:2830772.2830793, Nesbit:2004:AAD:1025127.1026003, Kumar:1998:ESL:279358.279404, Chen:2004:ACS:1072448.1072476, Cantin:2006:SP:1168857.1168892, Kim:2016:PCB:3195638.3195711, bingo_dpc3} divide the memory address space into fixed-size sections, named \componentname{Spatial Regions}, and learn the memory access patterns over these sections. The learned access patterns are then used for prefetching future memory references when the application touches the \emph{same} or \emph{similar} \componentname{Spatial Regions}. 

Spatial data prefetchers impose low area overhead because they store \emph{offsets} (i.e., the distance of a block address from the beginning of a \componentname{Spatial Region}) or \emph{deltas} (i.e., the distance of two consecutive accesses that fall into a \componentname{Spatial Region}) as their metadata information, and not complete addresses. Another equally remarkable strength of spatial data prefetchers is their ability to eliminate compulsory cache misses. Compulsory cache misses are a major source of performance degradation in important classes of applications, e.g., scan-dominated workloads, where scanning large volumes of data produces a bulk of unseen memory accesses that cannot be captured by caches~\cite{bakhshalipour2019bingo}. By utilizing the pattern that was observed in a past \componentname{Spatial Region} to a new unobserved \componentname{Spatial Region}, spatial prefetchers can alleviate the compulsory cache misses, significantly enhancing system performance.

The critical limitation of spatial data prefetching is its ineptitude in predicting pointer-chasing--caused cache misses. As dynamic objects can potentially be allocated everywhere in the memory, pointer-chasing accesses do not necessarily exhibit spatial correlation, producing bulks of dependent cache misses for which spatial prefetchers can do very little (cf.~Section~\ref{chapter:temporal_prefetching}).

We include two state-of-the-art spatial prefetching techniques: (1) \methodname{Spatial Memory Streaming}~\cite{Somogyi:2006:SMS:1135775.1136508}, and (2) \methodname{Variable Length Delta Prefetcher}~\cite{Shevgoor:2015:EPC:2830772.2830793}.

\subsection{\methodname{Spatial Memory Streaming (SMS)}}

\label{chapter:spatial_prefetching:sms}

\methodname{SMS} is a state-of-the-art spatial prefetcher that was proposed and evaluated in the context of server and scientific applications. Whenever a \componentname{Spatial Region} is requested for the first time, \methodname{SMS} starts to observe and record accesses to that \componentname{Spatial Region} as long as the \componentname{Spatial Region} is actively used by the application. Whenever the \componentname{Spatial Region} is no longer utilized (i.e., the corresponding blocks of the \componentname{Spatial Region} start to be evicted from the cache), \methodname{SMS} stores the information of the observed accesses in its metadata table, named \componentname{Pattern History Table (PHT)}. 

\begin{figure}[h]
   \centering
   \includegraphics[width=0.7\textwidth]{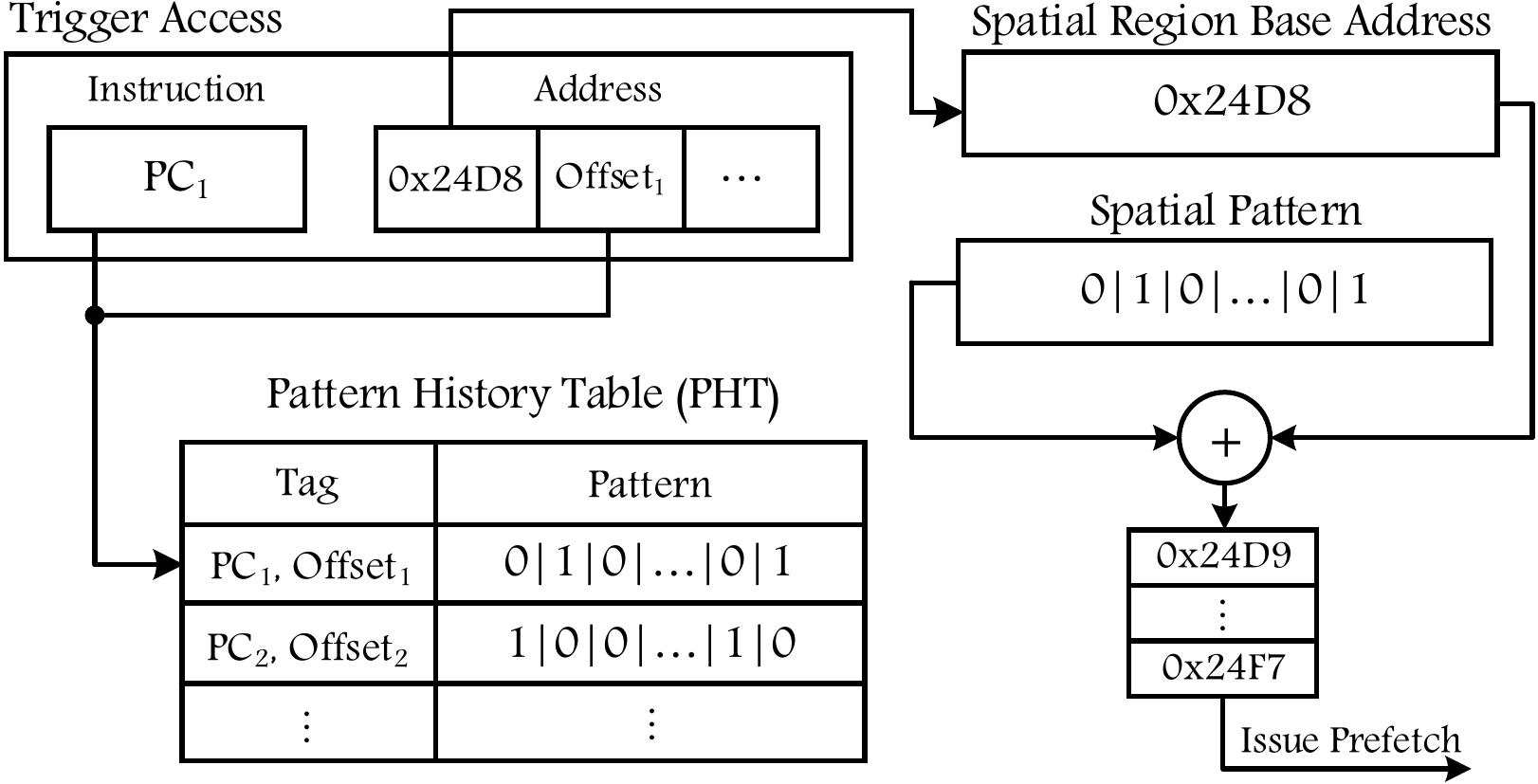}
   \caption{The organization of Spatial Memory Streaming (SMS).
   \label{fig:sms-diagram}}
\end{figure}

The information in \componentname{PHT} is stored in the form of $\langle{}event$, $pattern\rangle{}$. The $event$ is a piece of information to which the observed access pattern is correlated. That is, it is expected for the stored access pattern to be used whenever $event$ reoccurs in the future. \methodname{SMS} empirically chooses \componentname{PC+Offset} of the trigger access (i.e., the PC of the instruction that first accesses the \componentname{Spatial Region} combined with the distance of the first requested cache block from the beginning of the \componentname{Spatial Region}) as the $event$ to which the access patterns are correlated. Doing so, whenever a \componentname{PC+Offset} is reoccurred, the correlated access pattern history is used for issuing prefetch requests. The $pattern$ is the history of accesses that happen in every \componentname{Spatial Region}. \methodname{SMS} encodes the patterns of the accesses as a \emph{bit vector}. In this manner, for every cache block in a \componentname{Spatial Region}, a bit is stored, indicating whether the block has been used during the latest usage of the \componentname{Spatial Region} (\texttt{`1'}) or not (\texttt{`0'}). Therefore, whenever a $pattern$ is going to be used, prefetch requests are issued only for blocks whose corresponding bit in the stored $pattern$ is \texttt{`1}.\texttt{'} Figure~\ref{fig:sms-diagram} shows the hardware realization of \methodname{SMS}.

\subsection{\methodname{Variable Length Delta Prefetcher (VLDP)}}

\methodname{VLDP} is a recent state-of-the-art spatial data prefetcher that relies on the similarity of \emph{delta} patterns among \componentname{Spatial Regions} of memory. \methodname{VLDP} records the distance between consecutive accesses that fall into \componentname{Spatial Regions} and uses them to predict future misses. The key innovation of \methodname{VLDP} is the deployment of \emph{multiple} prediction tables for predicting delta patterns. \methodname{VLDP} employs several history tables where each table keeps the metadata based on a specific length of the input history. 

\begin{figure}[h]
   \centering
   \includegraphics[width=0.7\textwidth]{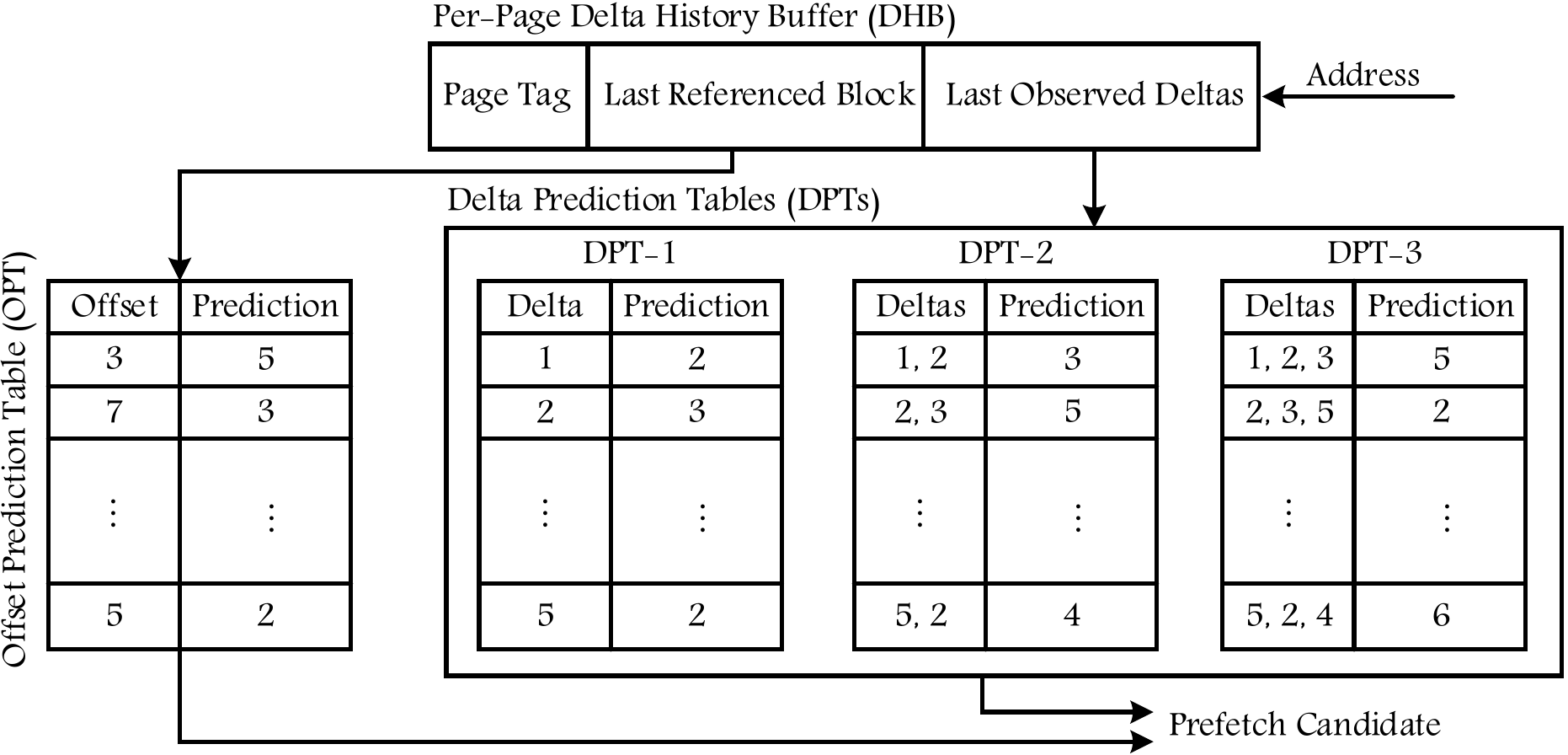}
   \caption{The organization of Variable Length Delta Prefetcher (VLDP).
   \label{fig:vldp-diagram}}
\end{figure}

Figure~\ref{fig:vldp-diagram} shows the metadata organization of \methodname{VLDP}. The three major components are \componentname{Delta History Buffer (DHB)}, \componentname{Delta Prediction Table (DPT)}, and \componentname{Offset Prediction Table (OPT)}. \componentname{DHB} is a small table that records the delta history of \emph{currently-active} \componentname{Spatial Regions}. Each entry in \componentname{DHB} is associated with an active \componentname{Spatial Region} and contains details like the \componentname{Last Referenced Block}. These details are used to index \componentname{OPT} and \componentname{DPTs} for issuing prefetch requests. 

\componentname{DPT} is a set of key-value pairs that correlates a delta sequence to the next expected delta. \methodname{VLDP} benefits from multiple \componentname{DPTs} where each \componentname{DPT} records the history with a different length of the input. \componentname{DPT}$-i$ associates a sequence of $i$ deltas to the next expected delta. For example, if the last three deltas in a \componentname{Spatial Region} are $d_{3}, d_{2},$ and $d_{1}$ ($d_{1}$ is the most recent delta), \componentname{DPT}-2 stores $[ \, \langle{}d_{3}, d_{2}\rangle{}\to{}d_{1}] \,$, while \componentname{DPT}-1 records $[ \, \langle{}d_{2}\rangle{}\to{}d_{1}]\,$. While looking up the \componentname{DPTs}, if several of them offer a prediction, the prediction of the table with the \emph{longest} sequence of deltas is used, because predictions that are made based on longer inputs are expected to be more accurate~\cite{bakhshalipour2019bingo}. This way, \methodname{VLDP} eliminates wrong predictions that are made by short inputs, enhancing both accuracy and miss coverage of the prefetcher. 

\componentname{OPT} is another metadata table of \methodname{VLDP}, that is indexed using the \emph{offset} (and not delta) of the first access to a \componentname{Spatial Region}. Merely relying on deltas for prefetching causes the prefetcher to need to observe at least first two accesses to a \componentname{Spatial Region} before issuing prefetch requests; however, there are many \emph{sparse} \componentname{Spatial Regions} in which a few, say, two, of the blocks are used by the application. Therefore, waiting for two accesses before starting the prefetching may divest the prefetcher of issuing enough prefetch requests when the application operates on a significant number of sparse \componentname{Spatial Regions}. Employing \componentname{OPT} enables \methodname{VLDP} to start prefetching immediately after the first access to \componentname{Spatial Regions}. \componentname{OPT} associates the offset of the first access of a \componentname{Spatial Region} to the next expected delta. After the first access to a \componentname{Spatial Region}, \componentname{OPT} is looked up using the offset of the access, and the output of the table is used for issuing a prefetch request. For the rest of the accesses to the \emph{Spatial Region} (i.e., second access onward), \methodname{VLDP} uses only \componentname{DPTs}.

Even though \methodname{VLDP} relies on prediction tables with a single next expected delta, it is still able to offer a prefetching lookahead larger than one (Section~\ref{chapter:introduction:background:lookahead}), using the proposed \emph{multi-degree} prefetching mechanism. In the \emph{multi-degree} mode, upon predicting the next delta in a \componentname{Spatial Region}, \methodname{VLDP} \emph{uses the prediction as an input} for \componentname{DPTs} to make more predictions.



\subsection{Conclusion}

Spatial prefetching has been proposed and developed to capture the similarity of access patterns among memory pages (e.g., if a program visits locations $\{A, B, C, D\}$ of Page $X$, it is probable that it visits locations $\{A, B, C, D\}$ of other pages as well). Spatial prefetching works because applications use data objects with a regular and fixed layout, and accesses reoccur when data structures are traversed. Spatial prefetching is appealing since it imposes low storage overhead to the system, paving the way for its adoption in the future systems.

\section{Temporal Prefetching}
\label{chapter:temporal_prefetching}

Temporal prefetching refers to replaying the sequence of past cache misses in order to avert future misses. Temporal data prefetchers~\cite{Joseph:1997:PUM:264107.264207, Chou:2007:LEC:1331699.1331727, Nesbit:2004:DCP:1072448.1072460, wenisch2009practical, Wenisch:2005:TSS:1069807.1069989, Solihin:2002:UUM:545215.545235, Jain:2013:LIM:2540708.2540730, bakhshalipour2018domino, bakhshalipour2017efficient} record the sequence of data misses in the order they appear and use the recorded history for predicting future data misses. Upon a new data miss, they search the history and find a matching entry and replay the sequence of data misses after the match in an attempt to eliminate potential future data misses. A tuned version of temporal prefetching has been implemented in \componentname{IBM Blue Gene/Q}, where it is called \methodname{List Prefetching}~\cite{haring2012ibm}.

Temporal prefetching is an ideal choice to eliminate long chains of \emph{dependent} cache misses, that are common in pointer-chasing applications (e.g., \emph{OLTP} and \emph{Web})~\cite{bakhshalipour2018domino}. A dependent cache miss refers to a memory operation that results in a cache miss and is dependent on data from a prior cache miss. Such misses have a marked effect on the execution performance of applications and impede the processor from making forward progress since both misses are fetched \emph{serially}~\cite{Hashemi:2016:ADC:3001136.3001184, bakhshalipour2018domino}. Because of the lack of strided/spatial correlation among dependent misses, stride and spatial prefetchers are usually unable to prefetch such misses~\cite{Somogyi:2009:SMS:1555754.1555766}; however, temporal prefetchers, by recording and replaying the sequences of data misses, can prefetch dependent cache misses and result in a significant performance improvement. 

Temporal prefetchers, on the other face of the coin, also have shortcomings. Temporal prefetching techniques exhibit low accuracy as they do not know where streams end. That is, in the foundation of temporal prefetching, there is no wealth of information about \emph{when prefetching should be stopped}; hence, temporal prefetchers continue issuing many prefetch requests until another triggering event occurs, resulting in a large overprediction. Moreover, as temporal prefetchers rely on address repetition, they are unable to prevent compulsory misses (unobserved misses) from happening. In other words, they can only prefetch cache misses that at least once have been observed in the past; however, there are many important applications (e.g., \emph{DSS}) in which the majority of cache misses occurs only once during the execution of the application~\cite{bakhshalipour2019bingo}, for which temporal prefetching can do nothing. Furthermore, as temporal prefetchers require to store the correlation between addresses, they usually impose large storage overhead (tens of megabytes) that cannot be accommodated on-the-chip next to the cores. Consequently, temporal prefetchers usually place their metadata tables off-the-chip in the main memory. Unfortunately, placing the history information off-the-chip increases the latency of accessing metadata, and more importantly, results in a drastic increase in the off-chip bandwidth consumption for fetching and updating the metadata. 

We include two state-of-the-art temporal prefetching techniques: (1) \methodname{Sampled Temporal Memory Streaming}~\cite{wenisch2009practical}, and (2) \methodname{Irregular Stream Buffer}~\cite{Jain:2013:LIM:2540708.2540730}.

\subsection{\methodname{Sampled Temporal Memory Streaming (STMS)}}
\label{chapter:temporal_prefetching:stms}

\methodname{STMS} is a state-of-the-art temporal data prefetcher that was proposed and evaluated in the context of server and scientific applications. The main observation behind \methodname{STMS} is that \emph{the length of temporal streams widely differs} across programs and across different streams in a particular program, as well; ranging from a couple to hundreds of thousands of cache misses. In order to efficiently store the information of various streams, \methodname{STMS} uses a circular FIFO buffer, named \componentname{History Table}, and appends every observed cache miss to its end. This way, the prefetcher is not required to fix a specific predefined length for temporal streams in the metadata organization, that would be resulted in wasting storage for streams shorten than the predefined length or discarding streams longer than it; instead, all streams are stored next to each other in a storage-efficient manner. For locating every address in the \componentname{History Table}, \methodname{STMS} uses an auxiliary set-associative structure, named \componentname{Index Table}. The \componentname{Index Table} stores a \emph{pointer} for every observed miss address to its last occurrence in the \componentname{History Table}. Therefore, whenever a cache miss occurs, the prefetcher first looks up the \componentname{Index Table} with the missed address and gets the corresponding pointer. Using the pointer, the prefetcher proceeds to the \componentname{History Table} and issues prefetch requests for addresses that have followed the missed address in the history. 

\begin{figure}[h]
   \centering
   \includegraphics[width=.7\textwidth]{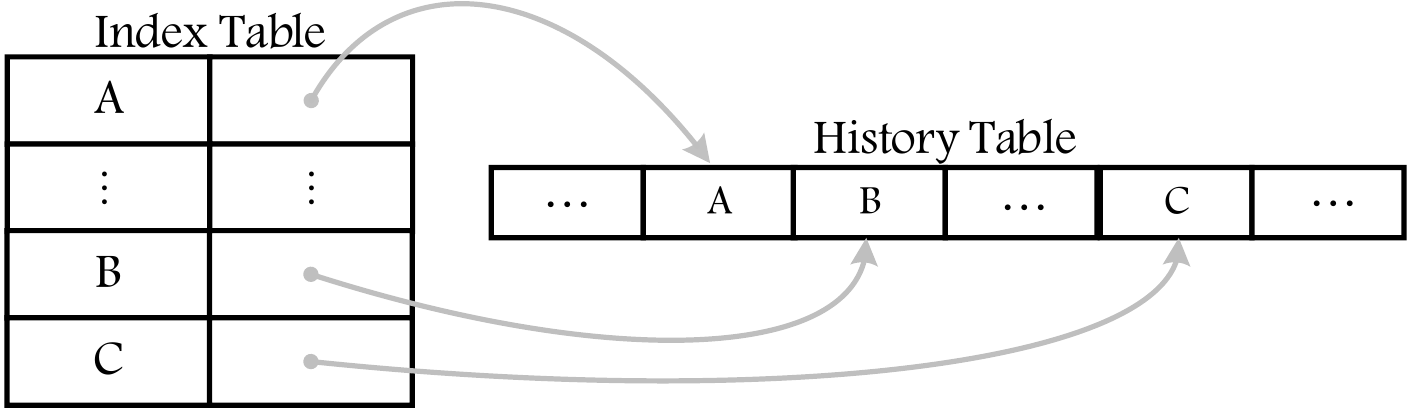}
   \caption{The organization of Sampled Temporal Memory Streaming (STMS).
   \label{fig:stms-diagram}}
\end{figure}

Figure~\ref{fig:stms-diagram} shows the metadata organization of \methodname{STMS}, which mainly consists of a \componentname{History Table} and an \componentname{Index Table}. As both tables require multi-megabyte storage for \methodname{STMS} to have reasonable miss coverage, both tables are placed off-the-chip in the main memory. Consequently, every access to these tables (read or update) should be sent to the main memory and brings/updates a cache block worth of data. That is, for every stream, \methodname{STMS} needs to wait for two long (serial) memory requests to be sent (one to read the \componentname{Index Table} and one to read the correct location of the \componentname{History Table}) and their responses to come back to the prefetcher before issuing prefetch requests for the stream. The delay of the two off-chip memory accesses, however, is compensated over several prefetch requests of a stream if the stream is long enough.

\subsection{\methodname{Irregular Stream Buffer (ISB)}}

\begin{figure}[h]
   \centering
   \includegraphics[width=.7\textwidth]{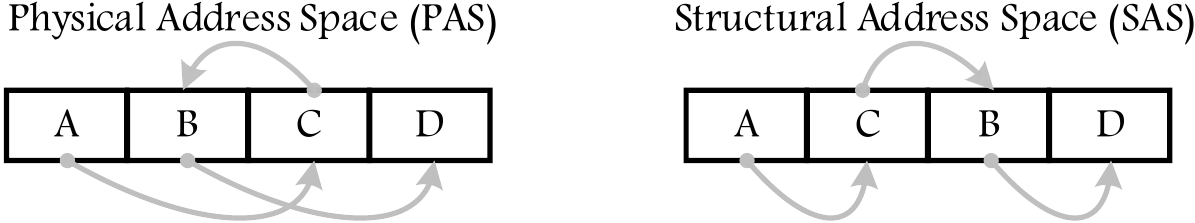}
   \caption{An example of linearizing scattered temporally-correlated memory references.
   \label{fig:isb-highlevel}}
\end{figure}

\methodname{ISB} is another state-of-the-art proposal for temporal data prefetching that targets irregular streams of temporally-correlated memory accesses. Unlike \methodname{STMS} that operates on the global miss sequences, \methodname{ISB} attempts to extract temporal correlation among memory references on a per load instruction basis (Section~\ref{chapter:introduction:ibsp}). The key innovation of \methodname{ISB} is the introduction of an extra \emph{indirection} level for storing metadata information. \methodname{ISB} defines a new conceptual address space, named \componentname{Structural Address Space (SAS)}, and \emph{maps} the temporally-correlated physical address to this address space in a way that they appear \emph{sequentially}. That is, with this indirection mechanism, physical addresses that are temporally-correlated and used one after another, regardless of their distribution in the \componentname{Physical Address Space (PAS)} of memory, become spatially-located and appear one after another in \componentname{SAS}. Figure~\ref{fig:isb-highlevel} shows a high-level example of this linearization. 

\begin{figure}[h]
   \centering
   \includegraphics[width=.7\textwidth]{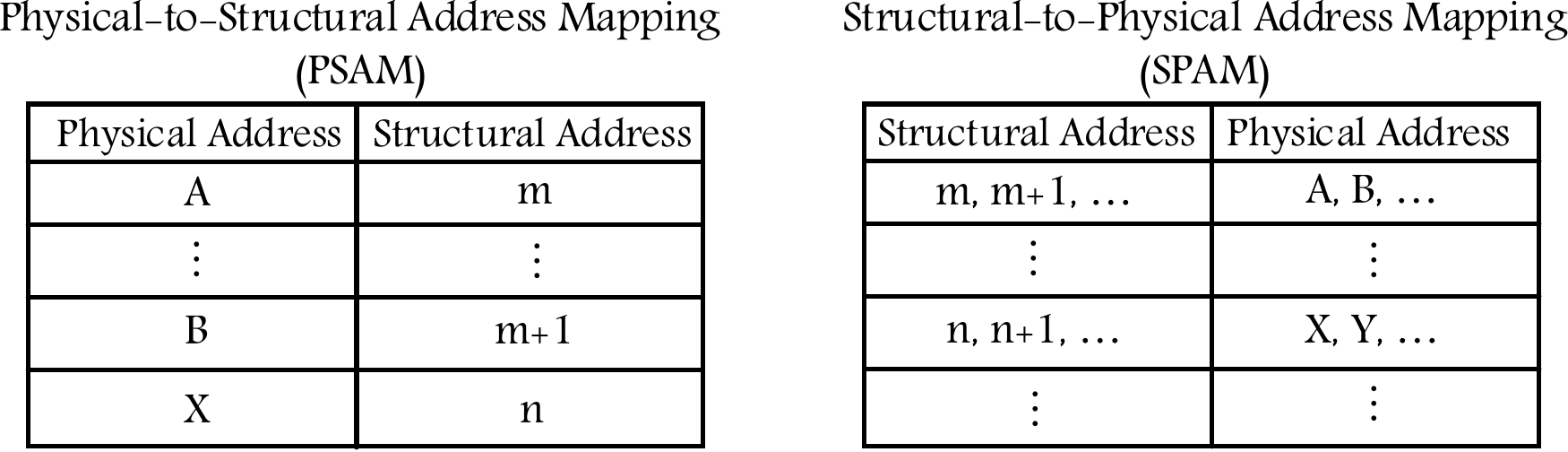}
   \caption{The organization of Irregular Stream Buffer (ISB). 
   \label{fig:isb-diagram}}
\end{figure}

\methodname{ISB} utilizes two tables to record a \emph{bidirectional} mapping between address in \componentname{PAS} and \componentname{SAS}: one table, named \componentname{Physical-to-Structural Address Mapping (PSAM)}, records temporally-correlated physical addresses and their mapping information (i.e., to which location in \componentname{SAS} they are mapped); the other table, named \componentname{Structural-to-Physical Address Mapping (SPAM)}, keeps the \emph{linearized} form of physical addresses in \componentname{SAS} and the corresponding mapping information (i.e., which physical addresses are mapped to every structural address). The main purpose of such a linearization is to represent the metadata in a spatially-located manner, paving the way to putting the metadata off-the-chip and \emph{caching} its content in on-chip structures~\cite{Burcea:2008:PV:1346281.1346301}. Like \methodname{STMS}, \methodname{ISB} puts its metadata information off-the-chip to save the precious SRAM storage; however, unlike them, \methodname{ISB} caches the content of its off-chip metadata tables in on-chip structures. Caching the metadata works for \methodname{ISB} as a result of the provided spatial locality. By caching the metadata information, \methodname{ISB} (1) provides faster access to metadata since the caches offer a high hit ratio, and it is not required to proceed to the off-chip memory for every metadata access, and (2) reduces the metadata-induced off-chip bandwidth overhead as many of the metadata manipulations coalesce in the on-chip caches. Figure~\ref{fig:isb-diagram} shows an overview of the metadata structures of \methodname{ISB}.

Another important contribution of \methodname{ISB} is the synchronization of off-chip metadata manipulations with Translation Lookaside Buffer (TLB) misses. That is, whenever a TLB miss occurs, concurrent with resolving the miss, \methodname{ISB} fetches the corresponding metadata information from the off-chip metadata tables; moreover, whenever a TLB entry is evicted, \methodname{ISB} evicts its corresponding entry from the on-chip metadata structures and updates the off-chip metadata tables. Doing so, \methodname{ISB} ensures that the required metadata is always present in the on-chip structures, significantly hiding the latency of off-chip memory accesses that would otherwise be exposed.

\subsection{Conclusion}
Temporal prefetching has been proposed and developed to capture temporally-correlated access patterns (i.e., the repetition of access patterns in the same order; e.g., if we observe $\{A, B, C, D\}$, then it is likely for $\{B, C, D\}$ to follow $\{A\}$ in the future). Temporal prefetching is well beneficial in the context of pointer-chasing applications, where applications produce bulks of cache misses that exhibit no spatial correlation, but temporal repetition. Temporal prefetchers, however, impose significant overheads to the system, which is still a grave concern in the research literature.

\section{Spatio-Temporal Data Prefetching}

Temporal and spatial prefetching techniques capture separate subsets of cache misses, and hence, each omits a considerable portion of cache misses unpredicted. As a considerable fraction of data misses is predictable only by one of the two prefetching techniques, spatio-temporal prefetching tries to combine them in order to reap the benefits of both methods. Another motivation for spatio-temporal prefetching is the fact that the effectiveness of temporal and spatial prefetching techniques varies across applications. As discussed, pointer-chasing application (e.g., \emph{OLTP}) produce long chains of dependent cache misses which cannot be effectively captured by spatial prefetching but temporal prefetching. On the contrary, scan-dominated applications (e.g., \emph{DSS}) produce a large number of compulsory cache misses that are predictable by spatial prefetchers and not temporal prefetchers. 

We include \methodname{Spatio-Temporal Memory Streaming (STeMS)}~\cite{Somogyi:2009:SMS:1555754.1555766}, as it is the only proposal in this class of prefetching techniques. 

\subsection{\methodname{Spatio-Temporal Memory Streaming (STeMS)}}
\label{chapter:spatio_temporal_prefetching:stems}

\methodname{STeMS} synergistically integrates spatial and temporal prefetching techniques in a unified prefetcher; \methodname{STeMS} uses a temporal prefetcher to capture the stream of \emph{trigger accesses} (i.e., the first access to each spatial region) and a spatial prefetcher to predict the expected misses \emph{within} the spatial regions. The metadata organization of \methodname{STeMS} mainly consists of the metadata tables of \methodname{STMS}~\cite{wenisch2009practical} and \methodname{SMS}~\cite{Somogyi:2006:SMS:1135775.1136508}. \methodname{STeMS}, however, seeks to stream the sequence of cache misses \emph{in the order they have been generated by the processor}, regardless of how the corresponding metadata information has been stored in the history tables of \methodname{STMS} and \methodname{SMS}. To do so, \methodname{STeMS} employs a \componentname{Reconstruction Buffer} which is responsible for reordering the prefetch requests generated by the temporal and the spatial prefetchers of \methodname{STeMS} so as to send prefetch requests (and deliver their responses) in the order the processor is supposed to consume them. 

\begin{figure}[h]
   \centering
   \includegraphics[width=0.7\textwidth]{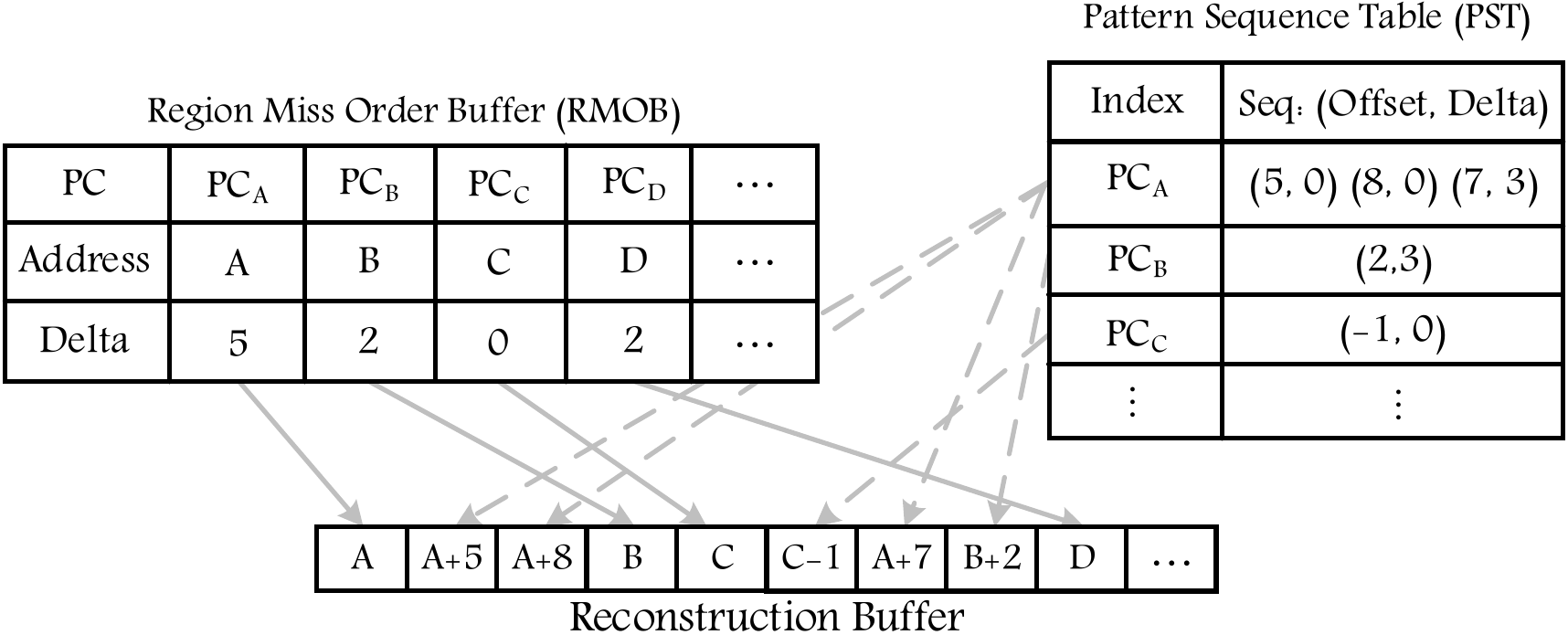}
   \caption{The organization of Spatio-Temporal Memory Streaming (STeMS) and the reconstruction process. 
   \label{fig:stems-diagram}}
\end{figure}

For enabling the \emph{reconstruction} process, the metadata tables of \methodname{SMS} and \methodname{STMS} are slightly modified. \methodname{SMS} is modified to record the order of the accessed cache blocks within a spatial region by encoding spatial patterns as ordered lists of offsets, stored in \componentname{Patterns Sequence Table (PST)}. Although \componentname{PST} is less compact than \componentname{PHT} (in the original \methodname{SMS}), the offset lists maintain the order required for accurately interleaving temporal and spatial streams. \methodname{STMS} is also modified and records only spatial triggers (and not all events as in \methodname{STMS}) in a \componentname{Region Miss Order Buffer (RMOB)}. Moreover, entries in both spatial and temporal streams are augmented with a \emph{delta} field. The delta field in a spatial (temporal) stream represents the number of events from the temporal (spatial) stream that is interleaved between the current and next events of the same type. Figure~\ref{fig:stems-diagram} gives an example of how \methodname{STeMS} reconstructs the total miss order.

\subsection{Conclusion}

Spatio-temporal prefetching has been proposed and developed to capture both temporal and spatial memory access patterns of applications. Spatio-temporal prefetching is based on the observation that temporal and spatial prefetchers each target a specific subset of cache misses, and leave the rest uncovered. Spatial-temporal data prefetching tries to synergistically capture both types of patterns, that any of the temporal or spatial prefetcher lonely cannot.

\section{State-of-The-Art Data Prefetcher}

In this chapter, we describe our own proposals for efficient data prefetching that have been published in recent years. We include them in the chronological order based on their publication date. 

\subsection{Domino Temporal Data Prefetcher}
\label{chapter:state-of-the-art-prefetchers:domino}

\begin{figure}[h]
   \centering
   \includegraphics[width=.7\textwidth]{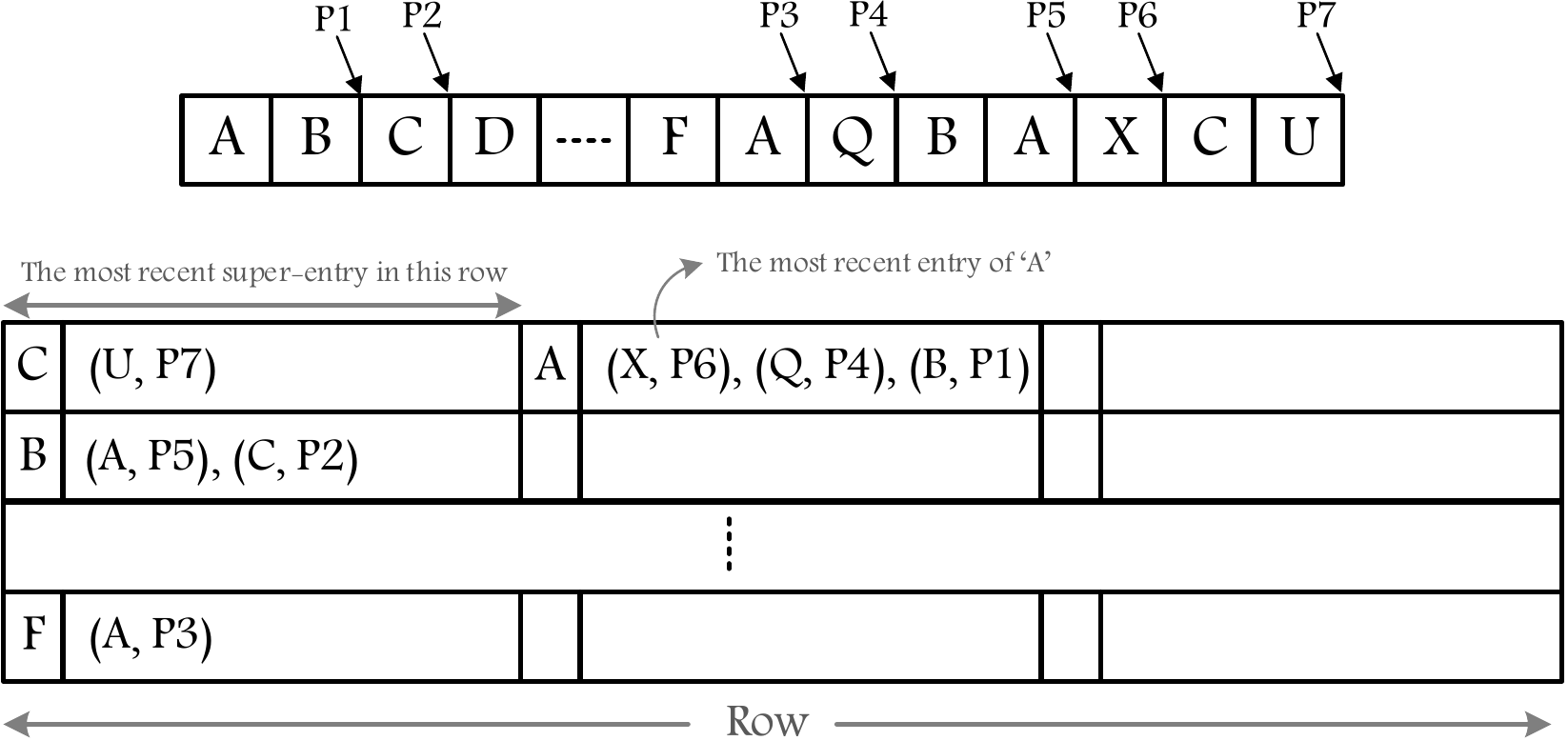}
   \caption{The details of the Enhanced Index Table in Domino prefetcher~\cite{bakhshalipour2018domino}.
   \label{fig:domino}}
\end{figure}

\methodname{Domino} is a state-of-the-art temporal data prefetcher that is built upon \methodname{STMS} (Section~\ref{chapter:temporal_prefetching:stms}) and seeks to improve its effectiveness. \methodname{Domino} is based on the observation that a single miss address, as used in the lookup mechanism of \methodname{STMS}, cannot always identify the correct miss stream in the history. Therefore, \methodname{Domino} provides a mechanism to look up the history of miss addresses with a combination of the last \emph{one or two} miss addresses. To do so, \methodname{Domino} replaces the \componentname{Index Table} of \methodname{STMS} with a novel structure, named \componentname{Enhanced Index Table (EIT)}. \componentname{EIT} like the \componentname{Index Table} of \methodname{STMS} stores a pointer for each address in the history; but unlike it, \emph{keeps the subsequent miss of each address}, additionally. Having the next miss of every address in the \componentname{EIT} enables \methodname{Domino} to find the correct stream in the history using the last one or two misses addresses. Moreover, with this organization, \methodname{Domino} becomes able to start prefetching (i.e., issuing the first prefetch request) right after touching \componentname{EIT}. That is, unlike \methodname{STMS} that needs to wait for two serial memory accesses (one for \componentname{Index Table}, then another for \componentname{History Table}) to start prefetching, \methodname{Domino} can start prefetching immediately after accessing \componentname{EIT}, because \componentname{EIT} contains the address of the first prefetch candidate. Starting prefetching sooner, causes \methodname{Domino} to offer superior timeliness as compared to \methodname{STMS}. Figure~\ref{fig:domino} shows the organization of the \componentname{EIT}.

The \componentname{EIT} is indexed by a \emph{single} miss address. Associated with every tag, there are several address-pointer
pairs, where the address is a miss of the core and the pointer is a location in the \componentname{History Table}. An \textit{(a, p)} pair associated to tag \textit{t} indicates that the pointer to the last occurrence of miss address \textit{t} followed by
\textit{a} is \textit{p}. The tag along with its associated address-pointer pairs is called a \emph{super-entry}, and every address-pointer pair is named an \emph{entry}. Every \emph{row} of the \componentname{EIT} has several \emph{super-entries}, and each \emph{super-entry} has several \emph{entries}. \methodname{Domino} keeps the LRU stack among both the \emph{super-entries} and the \emph{entries} within each \emph{super-entry}. Upon a cache miss, \methodname{Domino} uses the missed address to fetch a \emph{row} of the \componentname{EIT}. Then, \methodname{Domino} attempts to find the \emph{super-entry} associated with the missed address. In case a match is not found, nothing will be done, and otherwise, a prefetch will be sent for the address field of the most recent \emph{entry} in the found \emph{super-entry}. When the next triggering event occurs (miss or prefetch hit), \methodname{Domino} searches the \emph{super-entry} and picks the \emph{entry} for which the address field matches the triggering event. In case no match is found, \methodname{Domino} uses the triggering event to bring another \emph{row} from the \componentname{EIT}. Otherwise, \methodname{Domino} sends a request to read the \emph{row} of the \componentname{History Table} pointed to by the pointer field of the matched \emph{entry}. Once the sequence of miss addresses from the \componentname{History Table} arrives, \methodname{Domino} issues prefetch requests.

\subsection{Bingo Spatial Data Prefetcher}
\label{chapter:state-of-the-art-prefetchers:bingo}

\methodname{Bingo} is a recent proposal for spatial data prefetching, as well as the runner-up (and the winner in the multi-core evaluations) of The Third Data Prefetching Championship (DPC-3)~\cite{dpc3}. \methodname{Bingo} is based on the observation that assigning footprint information to a single event is suboptimal as compared to a case where footprints are correlated with \emph{multiple} events. \methodname{Bingo} discusses that events either are \emph{short}, that while their probability of recurrence is high, assigning footprints to them results in low accuracy, or are \emph{long}, that while prefetching using them results in high accuracy, much of the opportunity gets lost since the probability of recurring them is quite low. \methodname{Bingo}, for this reason, proposes to associate the observed footprint information to multiple events in order to provide both high opportunity and high accuracy. More specifically, \methodname{Bingo} correlates the observed footprint information of various pages with both \componentname{PC+Address} and \componentname{PC+Offset} of trigger accesses. In the context of \methodname{Bingo}, \componentname{PC+Address} is considered as a long event, while \componentname{PC+Offset} is treated as a short event. Whenever the time for prefetching comes (i.e., a triggering access occurs), \methodname{Bingo} uses the footprint that is associated with the longest occurred event for prefetching (i.e., \componentname{PC+Address}; and, if no history is recorded for \componentname{PC+Address}, \componentname{PC+Offset}). 

A naive implementation of \methodname{Bingo} requires two distinct \componentname{PHTs}: one table maintains the history of footprints observed after each \componentname{PC+Address}, while the other keeps the footprint metadata associated with \componentname{PC+Offset}. Upon looking for a pattern to prefetch, logically, first, the \componentname{PC+Address} of the trigger access is used to search the long \componentname{PHT}. If a match is found, the corresponding footprint is utilized to issue prefetch requests. Otherwise, the \componentname{PC+Offset} of the trigger access is used to look up the short \componentname{PHT}. In case of a match, the footprint metadata of the matched entry will be used for prefetching. If no matching entry is found, no prefetch will be issued. Such an implementation, however, would impose significant storage overhead. Authors in \methodname{Bingo} observe that, in the context of spatial data prefetching, a considerable fraction of the metadata that is stored in the \componentname{PHTs} are \emph{redundant}. That is, there are many cases where both metadata tables (tables associated with long and short events) offer the same prediction.

\begin{figure}[h]
   \centering
   \includegraphics[width=0.7\textwidth]{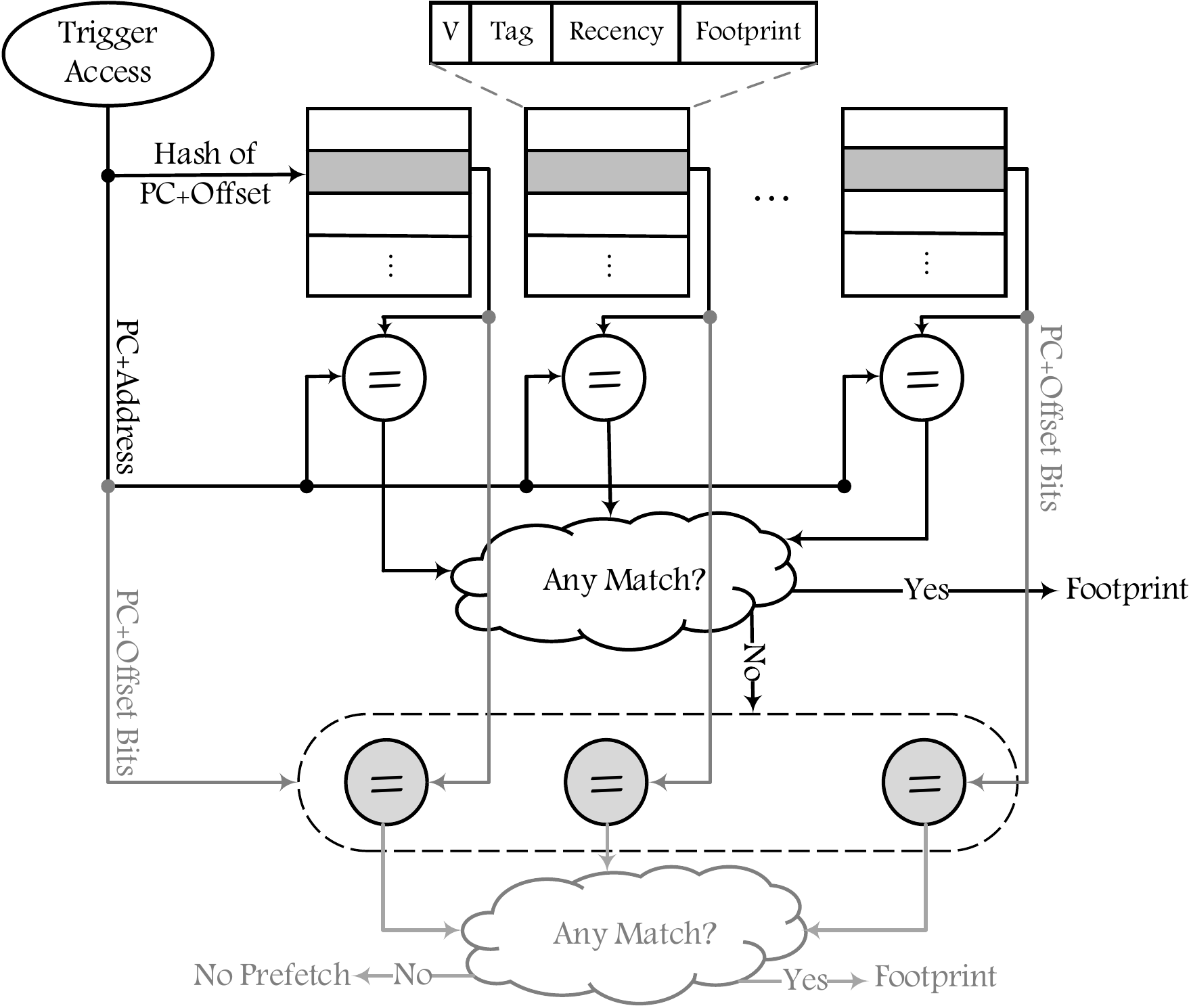}
   \caption{The details of the PHT lookup in \methodname{Bingo} prefetcher. Gray parts indicate the case where lookup with long event fails to find a match. Each large rectangle indicates a physical way of the history table~\cite{bakhshalipour2019bingo}.
   \label{fig:bingo}}
\end{figure}

To efficiently eliminate redundancies in the metadata storage, instead of using multiple history tables, \methodname{Bingo} proposes \emph{having a single history table but looking it up multiple times, each time with a different event}. Figure~\ref{fig:bingo} details the practical design of \methodname{Bingo} which uses only one \componentname{PHT}. The main insight is that \emph{short events are carried in long events}. That is, by having the long event at hand, one can find out what the short events are, just by ignoring parts of the long event. For the case of \methodname{Bingo}, the information of \componentname{PC+Offset} is carried in \componentname{PC+Address}; therefore, by knowing the \componentname{PC+Address}, the \componentname{PC+Offset} is also known. To exploit this phenomenon, \methodname{Bingo} proposes having only one history table which \emph{stores just the history of the long event but is looked up with both long and short events}. For the case of \methodname{Bingo}, the history table stores footprints which were observed after each \componentname{PC+Address} event, but is looked up with both the \componentname{PC+Address} and \componentname{PC+Offset} of the trigger access in order to offer high accuracy while not losing prefetching opportunities. 

To enable this, \methodname{Bingo} \emph{indexes the table with a hash of the shortest event} but \emph{tags it with the longest event}. Whenever a piece of information is going to be stored in the history metadata, it is associated with the longest event, and then stored in the history table. To do so, \emph{the bits corresponding to the shortest event are used for indexing the history table} to find the set in which the metadata should be stored; however, \emph{all bits of the longest event are used to tag the entry}. More specifically, with \methodname{Bingo}, whenever a new footprint is going to be stored in the metadata organization, it is associated with the corresponding \componentname{PC+Address}. To find a location in the history table for the new entry, a hash of only \componentname{PC+Offset} is used to index the table. By knowing the set, the baseline replacement algorithm (e.g., LRU) is used to choose a victim to open room for storing the new entry. After determining the location, the entry is stored in the history table, but all bits of the \componentname{PC+Address} are used for tagging the entry. 

Whenever there is a need for prediction, the \componentname{PHT} is first looked up with the longest event; if a match is found, it will be used to make a prediction. Otherwise, the table should be looked up with the next-longest event. As both long and short events are mapped to the same set, there is no need to check a new set; instead, the entries of the same set are tested to find a match with the shorter event. With \methodname{Bingo}, the table is first looked up with the \componentname{PC+Address} of the trigger access. If a match is found, the corresponding footprint metadata will be used for issuing prefetch requests. Otherwise, the table should be looked up with the \componentname{PC+Offset} of the trigger access. As both \componentname{PC+Address} and \componentname{PC+Offset} are mapped to the same set, there is no need to check a new set. That is, all the corresponding \componentname{PC+Offset} entries \emph{should be in the same set}. Therefore, the entries of the same set are tested to find a match. In this case, however, not all bits of the stored tags in the entries are necessary to match; only the \componentname{PC+Offset} bits need to be matched. This way, \methodname{Bingo} associates each footprint with more than one event (i.e., both \componentname{PC+Address} and \componentname{PC+Offset}) but store the footprint metadata in the table with only one of them (the longer one) to reduce the storage requirement. Doing so, redundancies are automatically eliminated because a metadata footprint is stored once with its \componentname{PC+Address} tag. 

\subsection{Multi-Lookahead Offset Prefetcher}
\methodname{Multi-Lookahead Offset Prefetcher (MLOP)}~\cite{mlop_dpc3} is a state-of-the-art \emph{offset prefetching}. Offset prefetching, in fact, is an evolution of stride prefetching (Section~\ref{chapter:introduction:ibsp}), in which, the prefetcher does \emph{not} try to detect strided streams. Instead, whenever a core requests for a cache block (e.g., $A$), the offset prefetcher prefetches the cache block that is distanced by $k$ cache lines (e.g., $A + k$), where $k$ is the \emph{prefetch offset}. In other words, offset prefetchers do not correlate the accessed address to any specific stream; rather, they treat the addresses individually, and based on the prefetch offset, they issue a prefetch request for every accessed address. Offset prefetchers have been shown to offer significant performance benefits while imposing small storage and logic overheads~\cite{sandbox, bop}.

The initial proposal for offset prefetching, named \methodname{Sandbox Prefetcher} (\methodname{SP})~\cite{sandbox}, attempts to find offsets that yield \emph{accurate} prefetch requests. To find such offsets, \methodname{SP} evaluates the prefetching accuracy of several predefined offsets (e.g., $-8, -7, \ldots{}, +8$) and finally allows offsets whose prefetching accuracy are beyond a certain threshold to issue actual prefetch requests. The later work, named \methodname{Best-Offset Prefetcher} (\methodname{BOP})~\cite{bop} tweaks \methodname{SP} and sets the \emph{timeliness} as the evaluation metric. \methodname{BOP} is based on the insight that accurate but \emph{late} prefetch requests do not accelerate the execution of applications as much as timely requests do. Therefore, \methodname{BOP} finds offsets that yield timely prefetch requests in an attempt to have the prefetched blocks ready before the processor actually asks for them. 

\methodname{MLOP} takes another step and proposes a novel offset prefetcher. \methodname{MLOP} is based on the observation that while \methodname{BOP} is able to generate timely prefetch requests, it loses much opportunity at covering cache misses because of \emph{relying on a {single} best offset and discarding many other appropriate offsets}. \methodname{BOP} evaluates several offsets and considers the offset that can generate the most timely prefetch requests as the best offset; then, it relies \emph{only} on this best offset to issue prefetch requests until another offset becomes better, and hence, the new best. In fact, this is a binary classification: the prefetch offsets are considered either as \emph{timely} offsets or \emph{late} offsets. After classification, the prefetcher does \emph{not} allow the so-called late offsets to issue any prefetch requests. However, there might be many other appropriate offsets that are less timely but are of value in that they can hide a significant fraction of cache miss delays.

To overcome the deficiencies of prior work, \methodname{MLOP} proposes to have a \emph{spectrum of timelinesses} for various prefetch offsets during their evaluations, rather than binarily classifying them. During the evaluation of various prefetch offsets, \methodname{MLOP} considers \emph{multiple lookaheads} for every prefetch offset: \emph{with which lookahead can an offset cover a specific cache miss?} To implement this, \methodname{MLOP} considers several lookaheads for each offset, and assigns score values to every offset with every lookahead, \emph{individually}. Finally, when the time for prefetching comes, \methodname{MLOP} finds the \emph{best offset for each lookahead} and allows it to issue prefetch requests; however, the prefetch requests for smaller lookaheads are \emph{prioritized} and issued first. By doing so, \methodname{MLOP} ensures that it allows the prefetcher to issue enough prefetch requests (i.e., various prefetch offsets are utilized; high miss coverage) while the timeliness is well considered (i.e., the prefetch requests are ordered). 

\begin{figure}[h]
 \centering
 \includegraphics[width=.6\textwidth]{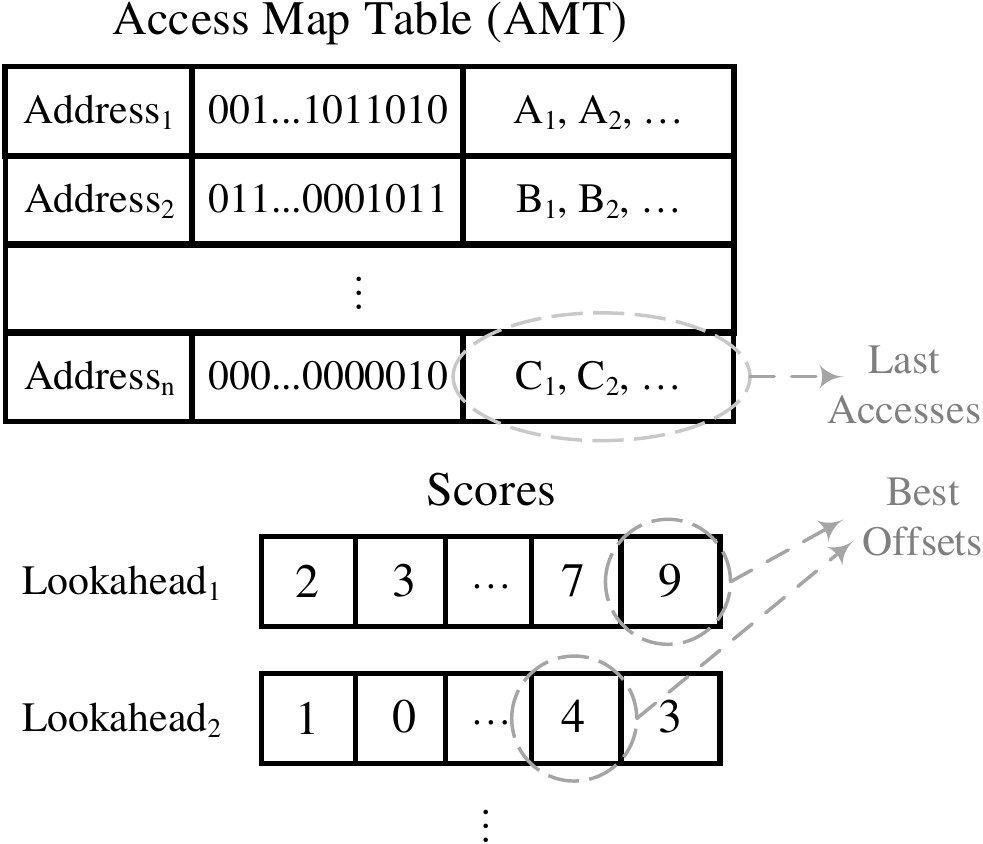}
 \caption{The hardware realization of our MLOP.
 \label{fig:mlop-design}}
\end{figure}

Figure~\ref{fig:mlop-design} shows an overview of \methodname{MLOP}. To extract offsets from access patterns, \methodname{MLOP} uses an \componentname{Access Map Table (AMT)}. The \componentname{AMT} keeps track of several recently-accessed addresses, along with a bit-vector for each base address. Each bit in the bit-vector corresponds to a cache block in the neighborhood of the address, indicating whether or not the block has been accessed. 

\methodname{MLOP} considers an evaluation period in which it evaluates several prefetch offsets and chooses the qualified ones for issuing prefetch requests later on. For every offset, it considers multiple levels of score where each level corresponds to a specific lookahead. The score of an offset at lookahead level $X$ indicates the number of cases where the offset prefetcher could have prefetched an access, at least $X$ accesses prior to occurrence. For example, the score of offsets at the lookahead level 1 indicates the number of cases where the offset prefetcher could have prefetched any of the futures accesses. 

To efficiently mitigate the negative effect of all predictable cache misses, \methodname{MLOP} selects \emph{one best offset from each lookahead level}. Then, during the actual prefetching, it allows all selected best offsets to issue prefetch requests. Doing so, \methodname{MLOP} ensures that it chooses enough prefetch offsets (i.e., does not suppress many qualified offsets like prior work~\cite{bop}), and will cover a significant fraction of cache misses that are predictable by offset prefetching. To handle the timeliness issue, \methodname{MLOP} ties to send the prefetch requests in a way that the application would have sent if there had not been any prefetcher: \methodname{MLOP} starts from lookahead level 1 (i.e., the accesses that are expected to happen the soonest) and issues the corresponding prefetch requests (using its best offset), then goes to the upper level; this process repeats. With this prioritization, \methodname{MLOP} tries to hide the latency of all predictable cache misses, as much as possible.

\subsection{Runahead Metadata}

\methodname{Runahead MetaData (RMD)}~\cite{rmd} is a general technique for harnessing \emph{pairwise-correlating} prefetching. Pairwise-correlating prefetching refers to methods that correlate every address (or more generally, every \emph{event}~\cite{bakhshalipour2019bingo}) with a \emph{single next prediction}. The next prediction can be the next expected address, with address-based pairwise-correlating prefetchers~\cite{Joseph:1997:PUM:264107.264207, hu2002timekeeping, hu:tcp, lai:deadblock, bakhshalipour2017efficient} or the next expected delta, with delta-based pairwise-correlating prefetchers~\cite{Kim:2016:PCB:3195638.3195711, Shevgoor:2015:EPC:2830772.2830793, bop}. 

A main challenge in pairwise-correlating prefetching is harnessing prefetching degree. Unlike streaming prefetchers~\cite{Wenisch:2005:TSS:1069807.1069989, wenisch2009practical, bakhshalipour2018domino} that prefetch multiple data addresses that follow the correlated address in the FIFO history buffer, or footprint-based prefetchers~\cite{Kumar:1998:ESL:279358.279404, Somogyi:2006:SMS:1135775.1136508, bakhshalipour2019bingo} that prefetch multiple data addresses whose corresponding bit in the bit-vector is set, pairwise-correlating prefetchers are limited to a single prediction per correlation entry; they cannot trivially issue \emph{multi-degree} prefetching. With this lookahead limitation, pairwise-correlating prefetching faces \emph{timeliness} as a major problem, in that, issuing merely a single prefetch request every time may not result in prefetch requests that cover the \emph{whole} latency of cache misses (Section~\ref{chapter:introduction:background}). 

What is typically employed in state-of-the-art pairwise-correlating data prefetchers as the de facto mechanism, including delta-based~\cite{Shevgoor:2015:EPC:2830772.2830793} and address-based~\cite{bakhshalipour2017efficient} ones, and even similar instruction prefetchers~\cite{spracklen2005effective}, is \emph{using the prediction as input to the metadata tables to make more predictions}: whenever a prediction is made, the prefetcher assumes it a correct prediction, and repeatedly indexes the metadata table with the prediction to make more predictions. While this approach has no storage overhead, it offers poor accuracy, as explicitly shown by recent work~\cite{Kim:2016:PCB:3195638.3195711, bakhshalipour2019bingo, bakhshalipour2018domino, spracklen2005effective}. The problem with such an approach is that the prefetcher has no information about \emph{how many times it should repeat this process}. In fact, this emanates from dissimilar stream lengths: if the prefetcher repeats this process $N$ times, for streams whose length is smaller than $N$, say $M$, it overprefetches $N-M$ addresses, resulting in inaccuracy; for streams longer than $N$, it may lose timeliness. Prior approaches that perform multi-degree prefetching in such a way, typically choose the degree of prefetching empirically based on a set of studied workloads. For example, Shevgoor et al.~\cite{Shevgoor:2015:EPC:2830772.2830793} set the degree to four; Bakhshalipour et al.~\cite{bakhshalipour2017efficient} set it to three. These numbers are chosen completely experimentally for a specific configuration and by examining a limited number of workloads, with which, the chosen number provides a reasonable trade-off between accuracy and timeliness. Obviously, limiting the degree to a certain predefined number neither is a solution that scales to various configurations and workloads, nor is optimal (w.r.t accuracy and timeliness) for the examined very configuration/workloads.

\methodname{RMD} proposes a novel solution to harness the multi-degree prefetching in the context of pairwise-correlating prefetchers. The key idea is to \emph{have separate metadata information for predicting the {next but one} expected event (e.g., the delta following the next delta; two deltas away from now)}. This way, in fact, \methodname{RMD} employs two separate metadata tables: one predicts the next event (\componentname{Distance1}; \componentname{D1}), the other predicts the next but one event (\componentname{Distance2}; \componentname{D2}), which is called \emph{Runahead Metadata Table}. When issuing multi-degree prefetching, the first prefetching is issued using only \componentname{D1}. For issuing the second prefetch, \componentname{D1} is searched using its previous prediction, similar to multi-degree prefetching of previous methods; meanwhile, \componentname{D2} is searched using the actual input (not prediction); the prefetch request is issued only if the prediction of both tables match; otherwise, the prefetching is finished. From the third prefetch request (if any) onward, both tables are searched using the corresponding inputs from the previous steps; if their predictions match, the prefetch request is issued and the process continues; otherwise, the prefetching is finished, concluding that the stream has come to an end. 

The reason for adding \componentname{D2} is to \emph{harness} the multi-degree prefetching of \componentname{D1}: until when the recursive lookups should resume? As \componentname{D2} operates one step ahead of \componentname{D1}, what \componentname{D2} offers is what \componentname{D1} is expected to offer in the next step. Hence, when \componentname{D1}'s second prediction (i.e., prediction using the previous prediction as input) is \emph{not} equal with \componentname{D2}'s prediction, it is intuitively concluded that the stream has been finished, and no further prefetch request is issued for the current stream. However, as long as the predictions match, the prefetcher continues prefetching to provide efficient timeliness, while preserving accuracy.

\begin{figure}[h]
   \centering
   \includegraphics[width=0.6\textwidth]{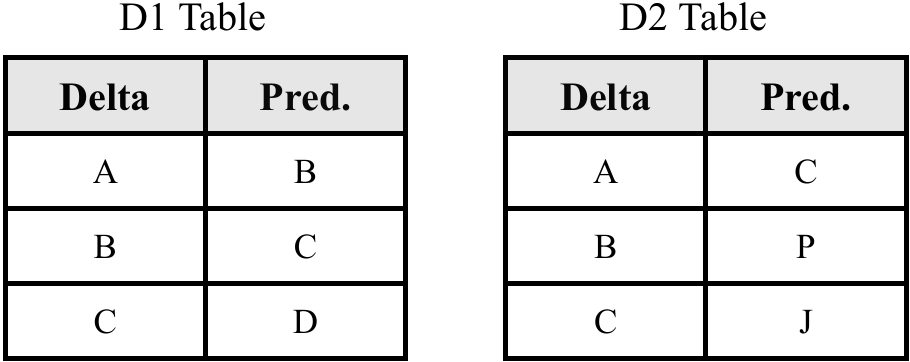} 
   \captionsetup{justification=centering}
   \caption{An illustration of how RMD works. 
   \label{fig:rmd}}
\end{figure}

Figure~\ref{fig:rmd} epitomizes how \emph{RMD} works. The entries in tables are interpreted in this way: $\langle{}A$, $B\rangle{}$ in \componentname{D1} shows that immediately after $A$, it is expected $B$ to happen; $\langle{}C$, $J\rangle{}$ in \componentname{D2} is intended to mean that two steps away from $C$, it is expected $J$ to happen. Consider that $A$ happens. \methodname{RMD} indexes \componentname{D1} by $A$. The prediction of \componentname{D1} is $B$; a prefetch request is issued for $B$. Then, \componentname{D1} is indexed by $B$; meanwhile, \componentname{D2} is indexed by $A$. The predictions of both \componentname{D1} and \componentname{D2} are $C$; their predictions match, and the prefetcher prefetches $C$. Then, \componentname{D1} is indexed by $C$ and \componentname{D2} by $B$. The prediction of \componentname{D1} is $D$, and the prediction of \componentname{D2} is $P$; their predictions do not match, and no further prefetch request is issued. 

\bibliographystyle{ieeetr}
\bibliography{ms}

\begin{thebibliography}{100}

\bibitem{Lotfi-Kamran:2012:SP:2337159.2337217}
P.~Lotfi-Kamran, B.~Grot, M.~Ferdman, S.~Volos, O.~Kocberber, J.~Picorel,
  A.~Adileh, D.~Jevdjic, S.~Idgunji, E.~Ozer, and B.~Falsafi, ``{Scale-Out
  Processors},'' in {\em Proceedings of the International Symposium on Computer
  Architecture (ISCA)}, pp.~500--511, 2012.

\bibitem{Grot:2012:ODT:2412372.2412724}
B.~Grot, D.~Hardy, P.~Lotfi-Kamran, C.~Nicopoulos, Y.~Sazeides, and B.~Falsafi,
  ``{Optimizing Data-Center TCO with Scale-Out Processors},'' {\em IEEE Micro},
  vol.~32, pp.~1--63, September 2012.

\bibitem{Lim:2008:UDN:1381306.1382148}
K.~Lim, P.~Ranganathan, J.~Chang, C.~Patel, T.~Mudge, and S.~Reinhardt,
  ``{Understanding and Designing New Server Architectures for Emerging
  Warehouse-Computing Environments},'' in {\em Proceedings of the International
  Symposium on Computer Architecture (ISCA)}, pp.~315--326, 2008.

\bibitem{Ferdman:2012:CCS:2150976.2150982}
M.~Ferdman, A.~Adileh, O.~Kocberber, S.~Volos, M.~Alisafaee, D.~Jevdjic,
  C.~Kaynak, A.~D. Popescu, A.~Ailamaki, and B.~Falsafi, ``{Clearing the
  Clouds: A Study of Emerging Scale-Out Workloads on Modern Hardware},'' in
  {\em Proceedings of the International Conference on Architectural Support for
  Programming Languages and Operating Systems (ASPLOS)}, pp.~37--48, 2012.

\bibitem{esmaili2018scale}
P.~Esmaili-Dokht, M.~Bakhshalipour, B.~Khodabandeloo, P.~Lotfi-Kamran, and
  H.~Sarbazi-Azad, ``{Scale-Out Processors \& Energy Efficiency},'' {\em arXiv
  preprint arXiv:1808.04864}, 2018.

\bibitem{ansari2019code}
A.~Ansari, P.~Lotfi-Kamran, and H.~Sarbazi-Azad, ``{Code Layout Optimization
  for Near-Ideal Instruction Cache},'' {\em IEEE Computer Architecture Letters
  (CAL)}, 2019.

\bibitem{mireshghallah2019energy}
F.~Mireshghallah, M.~Bakhshalipour, M.~Sadrosadati, and H.~Sarbazi-Azad,
  ``{Energy-Efficient Permanent Fault Tolerance in Hard Real-Time Systems},''
  {\em IEEE Transactions on Computers (TC)}, 2019.

\bibitem{jokar2020baldur}
M.~R. Jokar, J.~Qiu, F.~T. Chong, L.~L. Goddard, J.~M. Dallesasse, M.~Feng, and
  Y.~Li, ``{Baldur: A Power-Efficient and Scalable Network Using All-Optical
  Switches},'' in {\em Proceedings of the International Symposium on High
  Performance Computer Architecture (HPCA)}, pp.~153--166, 2020.

\bibitem{jokar2019high}
M.~R. Jokar, J.~Qiu, L.~L. Goddard, J.~M. Dallesasse, M.~Feng, Y.~Li, and F.~T.
  Chong, ``{A High-Performance and Energy-Efficient Optical Network Using
  Transistor Laser},'' in {\em TECHCON}, 2019.

\bibitem{al2008scalable}
M.~Al-Fares, A.~Loukissas, and A.~Vahdat, ``{A Scalable, Commodity Data Center
  Network Architecture},'' {\em ACM SIGCOMM Computer Communication Review
  (CCR)}, pp.~63--74, 2008.

\bibitem{farrington2010helios}
N.~Farrington, G.~Porter, S.~Radhakrishnan, H.~H. Bazzaz, V.~Subramanya,
  Y.~Fainman, G.~Papen, and A.~Vahdat, ``{Helios: A Hybrid Electrical/Optical
  Switch Architecture for Modular Data Centers},'' in {\em Proceedings of the
  ACM SIGCOMM Conference}, pp.~339--350, 2010.

\bibitem{Karkhanis:2004:FSP:1028176.1006729}
T.~S. Karkhanis and J.~E. Smith, ``{A First-Order Superscalar Processor
  Model},'' pp.~338--349, 2004.

\bibitem{Hameed:2010:USI:1815961.1815968}
R.~Hameed, W.~Qadeer, M.~Wachs, O.~Azizi, A.~Solomatnikov, B.~C. Lee,
  S.~Richardson, C.~Kozyrakis, and M.~Horowitz, ``{Understanding Sources of
  Inefficiency in General-Purpose Chips},'' in {\em Proceedings of the
  International Symposium on Computer Architecture (ISCA)}, pp.~37--47, ACM,
  2010.

\bibitem{Ferdman:2012:QME:2382553.2382557}
M.~Ferdman, A.~Adileh, O.~Kocberber, S.~Volos, M.~Alisafaee, D.~Jevdjic,
  C.~Kaynak, A.~D. Popescu, A.~Ailamaki, and B.~Falsafi, ``{Quantifying the
  Mismatch Between Emerging Scale-Out Applications and Modern Processors},''
  {\em ACM Transactions on Computer Systems (TOCS)}, vol.~30, pp.~15:1--15:24,
  November 2012.

\bibitem{vakil2018cache}
A.~Vakil-Ghahani, S.~Mahdizadeh-Shahri, M.-R. Lotfi-Namin, M.~Bakhshalipour,
  P.~Lotfi-Kamran, and H.~Sarbazi-Azad, ``{Cache Replacement Policy Based on
  Expected Hit Count},'' {\em IEEE Computer Architecture Letters (CAL)},
  vol.~17, no.~1, pp.~64--67, 2018.

\bibitem{ghahani2018making}
S.~A.~V. Ghahani, S.~M. Shahri, M.~Bakhshalipour, P.~Lotfi-Kamran, and
  H.~Sarbazi-Azad, ``{Making Belady-Inspired Replacement Policies More
  Effective Using Expected Hit Count},'' {\em arXiv preprint arXiv:1808.05024},
  2018.

\bibitem{Bakhshalipour:2019:EHD:3341324.3312740}
M.~Bakhshalipour, S.~Tabaeiaghdaei, P.~Lotfi-Kamran, and H.~Sarbazi-Azad,
  ``{Evaluation of Hardware Data Prefetchers on Server Processors},'' {\em ACM
  Computing Surveys (CSUR)}, vol.~52, pp.~52:1--52:29, June 2019.

\bibitem{bakhshalipour2019reducing}
M.~Bakhshalipour, A.~Faraji, S.~A.~V. Ghahani, F.~Samandi, P.~Lotfi-Kamran, and
  H.~Sarbazi-Azad, ``{Reducing Writebacks Through In-Cache Displacement},''
  {\em ACM Transactions on Design Automation of Electronic Systems (TODAES)},
  vol.~24, no.~2, p.~16, 2019.

\bibitem{livia}
E.~Lockerman, A.~Feldmann, M.~Bakhshalipour, A.~Stanescu, S.~Gupta, D.~Sanchez,
  and N.~Beckmann, ``{Livia: Data-Centric Computing Throughout the Memory
  Hierarchy},'' in {\em Proceedings of the International Conference on
  Architectural Support for Programming Languages and Operating Systems
  (ASPLOS)}, pp.~417--433, 2020.

\bibitem{esfeden2019corf}
H.~A. Esfeden, F.~Khorasani, H.~Jeon, D.~Wong, and N.~Abu-Ghazaleh, ``{CORF:
  Coalescing Operand Register File for GPUs},'' in {\em Proceedings of the
  International Conference on Architectural Support for Programming Languages
  and Operating Systems (ASPLOS)}, ACM, 2019.

\bibitem{khorasani2018register}
F.~Khorasani, H.~A. Esfeden, N.~Abu-Ghazaleh, and V.~Sarkar, ``{In-Register
  Parameter Caching for Dynamic Neural Nets with Virtual Persistent Processor
  Specialization},'' in {\em Proceedings of the International Symposium on
  Microarchitecture (MICRO)}, pp.~377--389, IEEE, 2018.

\bibitem{khorasani2018regmutex}
F.~Khorasani, H.~A. Esfeden, A.~Farmahini-Farahani, N.~Jayasena, and V.~Sarkar,
  ``{RegMutex: Inter-Warp GPU Register Time-Sharing},'' in {\em Proceedings of
  the International Symposium on Computer Architecture (ISCA)}, pp.~816--828,
  IEEE Press, 2018.

\bibitem{Kayaalp_RRI}
M.~Kayaalp, K.~N. Khasawneh, H.~A. Esfeden, J.~Elwell, N.~Abu-Ghazaleh,
  D.~Ponomarev, and A.~Jaleel, ``{RIC: Relaxed Inclusion Caches for Mitigating
  LLC Side-Channel Attacks},'' in {\em Proceedings of the Design Automation
  Conference (DAC)}, pp.~7:1--7:6, ACM, 2017.

\bibitem{hojabr2017customizing}
R.~Hojabr, M.~Modarressi, M.~Daneshtalab, A.~Yasoubi, and A.~Khonsari,
  ``{Customizing Clos Network-on-Chip for Neural Networks},'' {\em IEEE
  Transactions on Computers (TC)}, 2017.

\bibitem{knights_landing}
A.~Sodani, R.~Gramunt, J.~Corbal, H.-S. Kim, K.~Vinod, S.~Chinthamani,
  S.~Hutsell, R.~Agarwal, and Y.-C. Liu, ``{Knights Landing: Second-Generation
  Intel Xeon Phi Product},'' {\em IEEE Micro}, vol.~36, pp.~34--46, March 2016.

\bibitem{haring2012ibm}
R.~Haring, M.~Ohmacht, T.~Fox, M.~Gschwind, D.~Satterfield, K.~Sugavanam,
  P.~Coteus, P.~Heidelberger, M.~Blumrich, R.~Wisniewski, A.~Gara, G.~Chiu,
  P.~Boyle, N.~Chist, and C.~Kim, ``{The IBM Blue Gene/Q Compute Chip},'' {\em
  IEEE Micro}, vol.~32, no.~2, pp.~48--60, 2012.

\bibitem{amd_opteron}
P.~Conway and B.~Hughes, ``{The AMD Opteron Northbridge Architecture},'' {\em
  IEEE Micro}, vol.~27, pp.~10--21, March 2007.

\bibitem{horel1999ultrasparc}
T.~Horel and G.~Lauterbach, ``{UltraSPARC-III: Designing Third-Generation
  64-bit Performance},'' {\em IEEE Micro}, vol.~19, no.~3, pp.~73--85, 1999.

\bibitem{Wulf:1995:HMW:216585.216588}
W.~A. Wulf and S.~A. McKee, ``{Hitting the Memory Wall: Implications of the
  Obvious},'' {\em SIGARCH Comput. Archit. News}, vol.~23, pp.~20--24, March
  1995.

\bibitem{Trancoso:1997:MPD:548716.822671}
P.~Trancoso, J.-L. Larriba-Pey, Z.~Zhang, and J.~Torrellas, ``{The Memory
  Performance of DSS Commercial Workloads in Shared-Memory Multiprocessors},''
  in {\em Proceedings of the International Symposium on High Performance
  Computer Architecture (HPCA)}, pp.~250--260, 1997.

\bibitem{Ailamaki:1999:DMP:645925.671662}
A.~Ailamaki, D.~J. DeWitt, M.~D. Hill, and D.~A. Wood, ``{DBMSs on a Modern
  Processor: Where Does Time Go?},'' in {\em Proceedings of the International
  Conference on Very Large Data Bases (VLDB)}, pp.~266--277, 1999.

\bibitem{Hankins:2003:SCR:956417.956541}
R.~A. Hankins, T.~Diep, M.~Annavaram, B.~Hirano, H.~Eri, H.~Nueckel, and J.~P.
  Shen, ``{Scaling and Characterizing Database Workloads: Bridging the Gap
  Between Research and Practice},'' in {\em Proceedings of the International
  Symposium on Microarchitecture (MICRO)}, pp.~116--120, 2003.

\bibitem{hardavellas:database}
N.~Hardavellas, I.~Pandis, R.~Johnson, N.~G. Mancheril, A.~Ailamaki, and
  B.~Falsafi, ``{Database Servers on Chip Multiprocessors: Limitations and
  Opportunities},'' in {\em Proceedings of the Biennial Conference on
  Innovative Data Systems Research (CIDR)}, pp.~79--87, 2007.

\bibitem{bakhshalipour2018stacked}
M.~Bakhshalipour, H.~Zare, P.~Lotfi-Kamran, and H.~Sarbazi-Azad, ``{Die-Stacked
  DRAM: Memory, Cache, or MemCache?},'' {\em arXiv preprint arXiv:1809.08828},
  2018.

\bibitem{Rashidi:2018:IMP:3199680.3177965}
S.~Rashidi, M.~Jalili, and H.~Sarbazi-Azad, ``{Improving MLC PCM Performance
  Through Relaxed Write and Read for Intermediate Resistance Levels},'' {\em
  ACM Transactions on Architecture and Code Optimization (TACO)}, vol.~15,
  no.~1, pp.~12:1--12:31, 2018.

\bibitem{saeed_csur}
S.~Rashidi, M.~Jalili, and H.~Sarbazi-Azad, ``{A Survey on PCM Lifetime
  Enhancement Schemes},'' {\em ACM Computing Surveys (CSUR)}, 2019.

\bibitem{nesta_mirzaeian}
A.~Mirzaeian, H.~Homayoun, and A.~Sasan, ``{NESTA: Hamming Weight
  Compression-Based Neural Proc. Engine},'' in {\em The Proceedings of the Asia
  and South Pacific Design Automation Conference (ASPDAC)}, 2020.

\bibitem{tcd_mirzaeian}
A.~Mirzaeian, H.~Homayoun, and A.~Sasan, ``{TCD-NPE: A Re-Configurable and
  Efficient Neural Processing Engine, Powered by Novel Temporal-Carry-Deferring
  MACs},'' in {\em The Processings of the International Conference on
  ReConFigurable Computing and FPGAs (ReConFig)}, 2020.

\bibitem{armin_dsm}
S.~A. Vakil~Ghahani, M.~T. Kandemir, and J.~B. Kotra, ``{DSM: A Case for
  Hardware-Assisted Merging of DRAM Rows with Same Content},'' in {\em The
  Proceedings of the ACM on the Measurement and Analysis of Computing Systems
  (POMACS)}, 2020.

\bibitem{jeon2019locality}
H.~Jeon, H.~A. Esfeden, N.~B. Abu-Ghazaleh, D.~Wong, and S.~Elango,
  ``{Locality-aware GPU Register File},'' {\em IEEE Computer Architecture
  Letters (CAL)}, 2019.

\bibitem{jokar2018cooperative}
M.~R. Jokar, L.~Zhang, and F.~T. Chong, ``{Cooperative NV-NUMA: Prolonging
  Non-Volatile Memory Lifetime Through Bandwidth Sharing},'' in {\em
  Proceedings of the International Symposium on Memory Systems (MEMSYS)},
  pp.~67--78, 2018.

\bibitem{Nemirovsky:2013:MA:2502821}
M.~Nemirovsky and D.~M. Tullsen, {\em {Multithreading Architecture}}.
\newblock Morgan \& Claypool Publishers, 1st~ed., 2013.

\bibitem{Ryoo_OPA}
S.~Ryoo, C.~I. Rodrigues, S.~S. Baghsorkhi, S.~S. Stone, D.~B. Kirk, and
  W.-m.~W. Hwu, ``{Optimization Principles and Application Performance
  Evaluation of a Multithreaded GPU Using CUDA},'' in {\em Proceedings of the
  Symposium on Principles and Practice of Parallel Programming (PPoPP)},
  pp.~73--82, ACM, 2008.

\bibitem{akkary1998dynamic}
H.~Akkary and M.~A. Driscoll, ``{A Dynamic Multithreading Processor},'' in {\em
  Proceedings of the International Symposium on Microarchitecture (MICRO)},
  pp.~226--236, IEEE, 1998.

\bibitem{Cui:2010:SDM:1924943.1924958}
H.~Cui, J.~Wu, C.-C. Tsai, and J.~Yang, ``{Stable Deterministic Multithreading
  Through Schedule Memoization},'' in {\em Proceedings of the USENIX Conference
  on Operating Systems Design and Implementation (OSDI)}, pp.~207--221, USENIX
  Association, 2010.

\bibitem{bakhshalipour2018parallelizing}
M.~Bakhshalipour and H.~Sarbazi-Azad, ``{Parallelizing Bisection Root-Finding:
  A Case for Accelerating Serial Algorithms in Multicore Substrates},'' {\em
  arXiv preprint arXiv:1805.07269}, 2018.

\bibitem{Lo:1998:ADW:279358.279367}
J.~L. Lo, L.~A. Barroso, S.~J. Eggers, K.~Gharachorloo, H.~M. Levy, and S.~S.
  Parekh, ``{An Analysis of Database Workload Performance on Simultaneous
  Multithreaded Processors},'' in {\em Proceedings of the International
  Symposium on Computer Architecture (ISCA)}, pp.~39--50, 1998.

\bibitem{Collins:2001:SPL:379240.379248}
J.~D. Collins, H.~Wang, D.~M. Tullsen, C.~Hughes, Y.-F. Lee, D.~Lavery, and
  J.~P. Shen, ``{Speculative Precomputation: Long-Range Prefetching of
  Delinquent Loads},'' in {\em Proceedings of the International Symposium on
  Computer Architecture (ISCA)}, pp.~14--25, 2001.

\bibitem{Ganusov:2006:FEP:1187976.1187979}
I.~Ganusov and M.~Burtscher, ``{Future Execution: A Prefetching Mechanism That
  Uses Multiple Cores to Speed Up Single Threads},'' {\em ACM Transactions on
  Architecture and Code Optimization (TACO)}, vol.~3, pp.~424--449, December
  2006.

\bibitem{Lee:2009:PHT:1591896.1592265}
J.~Lee, C.~Jung, D.~Lim, and Y.~Solihin, ``{Prefetching with Helper Threads for
  Loosely Coupled Multiprocessor Systems},'' {\em IEEE Transactions on Parallel
  and Distributed Systems (TPDS)}, vol.~20, pp.~1309--1324, September 2009.

\bibitem{Kamruzzaman:2011:IPM:1950365.1950411}
M.~Kamruzzaman, S.~Swanson, and D.~M. Tullsen, ``{Inter-Core Prefetching for
  Multicore Processors Using Migrating Helper Threads},'' in {\em Proceedings
  of the International Conference on Architectural Support for Programming
  Languages and Operating Systems (ASPLOS)}, pp.~393--404, 2011.

\bibitem{Collins:2001:DSP:563998.564037}
J.~D. Collins, D.~M. Tullsen, H.~Wang, and J.~P. Shen, ``{Dynamic Speculative
  Precomputation},'' in {\em Proceedings of the International Symposium on
  Microarchitecture (MICRO)}, pp.~306--317, 2001.

\bibitem{Mutlu:2003:REA:822080.822823}
O.~Mutlu, J.~Stark, C.~Wilkerson, and Y.~N. Patt, ``{Runahead Execution: An
  Alternative to Very Large Instruction Windows for Out-of-Order Processors},''
  in {\em Proceedings of the International Symposium on High Performance
  Computer Architecture (HPCA)}, pp.~129--, 2003.

\bibitem{Mutlu:2005:TEP:1069807.1070000}
O.~Mutlu, H.~Kim, and Y.~N. Patt, ``{Techniques for Efficient Processing in
  Runahead Execution Engines},'' in {\em Proceedings of the International
  Symposium on Computer Architecture (ISCA)}, pp.~370--381, 2005.

\bibitem{Kadjo:2014:BBP:2742155.2742218}
D.~Kadjo, J.~Kim, P.~Sharma, R.~Panda, P.~Gratz, and D.~Jimenez, ``{B-Fetch:
  Branch Prediction Directed Prefetching for Chip-Multiprocessors},'' in {\em
  Proceedings of the International Symposium on Microarchitecture (MICRO)},
  pp.~623--634, 2014.

\bibitem{Hashemi:2016:ADC:3001136.3001184}
M.~Hashemi, Khubaib, E.~Ebrahimi, O.~Mutlu, and Y.~N. Patt, ``{Accelerating
  Dependent Cache Misses with an Enhanced Memory Controller},'' in {\em
  Proceedings of the International Symposium on Computer Architecture (ISCA)},
  pp.~444--455, 2016.

\bibitem{Ranganathan:1998:PDW:291069.291067}
P.~Ranganathan, K.~Gharachorloo, S.~V. Adve, and L.~A. Barroso, ``{Performance
  of Database Workloads on Shared-Memory Systems with Out-of-Order
  Processors},'' in {\em Proceedings of the International Conference on
  Architectural Support for Programming Languages and Operating Systems
  (ASPLOS)}, pp.~307--318, 1998.

\bibitem{bakhshalipour2018domino}
M.~Bakhshalipour, P.~Lotfi-Kamran, and H.~Sarbazi-Azad, ``{Domino Temporal Data
  Prefetcher},'' in {\em Proceedings of the International Symposium on
  High-Performance Computer Architecture (HPCA)}, pp.~131--142, IEEE, 2018.

\bibitem{Johnson:2007:SS:1325851.1325894}
R.~Johnson, S.~Harizopoulos, N.~Hardavellas, K.~Sabirli, I.~Pandis,
  A.~Ailamaki, N.~G. Mancheril, and B.~Falsafi, ``{To Share or Not to
  Share?},'' in {\em Proceedings of the International Conference on Very Large
  Data Bases (VLDB)}, pp.~351--362, 2007.

\bibitem{Larus:2002:UCE:647057.713864}
J.~R. Larus and M.~Parkes, ``{Using Cohort-Scheduling to Enhance Server
  Performance},'' in {\em Proceedings of the General Track of the Annual
  Conference on USENIX Annual Technical Conference (ATEC)}, pp.~103--114, 2002.

\bibitem{Zhang:2006:SPE:1121992.1122392}
W.~Zhang, B.~Calder, and D.~M. Tullsen, ``{A Self-Repairing Prefetcher in an
  Event-Driven Dynamic Optimization Framework},'' in {\em Proceedings of the
  International Symposium on Code Generation and Optimization (CGO)},
  pp.~50--64, 2006.

\bibitem{Luk:1996:CPR:237090.237190}
C.-K. Luk and T.~C. Mowry, ``{Compiler-Based Prefetching for Recursive Data
  Structures},'' in {\em Proceedings of the International Conference on
  Architectural Support for Programming Languages and Operating Systems
  (ASPLOS)}, pp.~222--233, 1996.

\bibitem{Roth:1999:EJP:300979.300989}
A.~Roth and G.~S. Sohi, ``{Effective Jump-Pointer Prefetching for Linked Data
  Structures},'' in {\em Proceedings of the International Symposium on Computer
  Architecture (ISCA)}, pp.~111--121, 1999.

\bibitem{Chilimbi:2002:DHD:512529.512554}
T.~M. Chilimbi and M.~Hirzel, ``{Dynamic Hot Data Stream Prefetching for
  General-Purpose Programs},'' in {\em Proceedings of the ACM SIGPLAN
  Conference on Programming Language Design and Implementation (PLDI)},
  pp.~199--209, 2002.

\bibitem{Chen:2007:IHJ:1272743.1272747}
S.~Chen, A.~Ailamaki, P.~B. Gibbons, and T.~C. Mowry, ``{Improving Hash Join
  Performance Through Prefetching},'' {\em ACM Transactions on Database Systems
  (TODS)}, vol.~32, August 2007.

\bibitem{Hughes:2005:MPL:1066486.1066491}
C.~J. Hughes and S.~V. Adve, ``{Memory-Side Prefetching for Linked Data
  Structures for Processor-in-Memory Systems},'' {\em Journal of Parallel and
  Distributed Computing}, vol.~65, pp.~448--463, April 2005.

\bibitem{Solihin:2002:UUM:545215.545235}
Y.~Solihin, J.~Lee, and J.~Torrellas, ``{Using a User-Level Memory Thread for
  Correlation Prefetching},'' in {\em Proceedings of the International
  Symposium on Computer Architecture (ISCA)}, pp.~171--182, 2002.

\bibitem{Yedlapalli:2013:MMI:2523721.2523761}
P.~Yedlapalli, J.~Kotra, E.~Kultursay, M.~Kandemir, C.~R. Das, and
  A.~Sivasubramaniam, ``{Meeting Midway: Improving CMP Performance with
  Memory-Side Prefetching},'' in {\em Proceedings of the International
  Conference on Parallel Architectures and Compilation Techniques (PACT)},
  pp.~289--298, 2013.

\bibitem{Mittal:2016:SRP:2966278.2907071}
S.~Mittal, ``{A Survey of Recent Prefetching Techniques for Processor
  Caches},'' {\em ACM Computing Surveys (CSUR)}, vol.~49, pp.~35:1--35:35,
  August 2016.

\bibitem{Srinath:2007:FDP:1317533.1318101}
S.~Srinath, O.~Mutlu, H.~Kim, and Y.~N. Patt, ``{Feedback Directed Prefetching:
  Improving the Performance and Bandwidth-Efficiency of Hardware
  Prefetchers},'' in {\em Proceedings of the International Symposium on High
  Performance Computer Architecture (HPCA)}, pp.~63--74, 2007.

\bibitem{Falsafi:2014:PHP:2643033}
B.~Falsafi and T.~F. Wenisch, {\em {A Primer on Hardware Prefetching}}.
\newblock Morgan \& Claypool Publishers, 2014.

\bibitem{Dahlgren:1995:EHS:527072.822612}
F.~Dahlgren and P.~Stenstrom, ``{Effectiveness of Hardware-Based Stride and
  Sequential Prefetching in Shared-memory Multiprocessors},'' in {\em
  Proceedings of the International Symposium on High Performance Computer
  Architecture (HPCA)}, pp.~68--, 1995.

\bibitem{Chilimbi:2001:STD:645988.674166}
T.~M. Chilimbi, ``{On the Stability of Temporal Data Reference Profiles},'' in
  {\em Proceedings of the International Conference on Parallel Architectures
  and Compilation Techniques (PACT)}, pp.~151--160, 2001.

\bibitem{bakhshalipour2019bingo}
M.~Bakhshalipour, M.~Shakerinava, P.~Lotfi-Kamran, and H.~Sarbazi-Azad,
  ``{Bingo Spatial Data Prefetcher},'' in {\em Proceedings of the International
  Symposium on High-Performance Computer Architecture (HPCA)}, pp.~399--411,
  2019.

\bibitem{Kim:2016:PCB:3195638.3195711}
J.~Kim, S.~H. Pugsley, P.~V. Gratz, A.~L.~N. Reddy, C.~Wilkerson, and
  Z.~Chishti, ``{Path Confidence Based Lookahead Prefetching},'' in {\em
  Proceedings of the International Symposium on Microarchitecture (MICRO)},
  pp.~60:1--60:12, 2016.

\bibitem{Mehta:2014:MCP:2597652.2597660}
S.~Mehta, Z.~Fang, A.~Zhai, and P.-C. Yew, ``{Multi-Stage Coordinated
  Prefetching for Present-Day Processors},'' in {\em Proceedings of the
  International Conference on Supercomputing (ICS)}, pp.~73--82, 2014.

\bibitem{bakhshalipour2018fast}
M.~Bakhshalipour, P.~Lotfi-Kamran, A.~Mazloumi, F.~Samandi, M.~Naderan-Tahan,
  M.~Modarressi, and H.~Sarbazi-Azad, ``{Fast Data Delivery for Many-Core
  Processors},'' {\em IEEE Transactions on Computers (TC)}, vol.~67, no.~10,
  pp.~1416--1429, 2018.

\bibitem{Ebrahimi:2009:CCM:1669112.1669154}
E.~Ebrahimi, O.~Mutlu, C.~J. Lee, and Y.~N. Patt, ``{Coordinated Control of
  Multiple Prefetchers in Multi-Core Systems},'' in {\em Proceedings of the
  International Symposium on Microarchitecture (MICRO)}, pp.~316--326, 2009.

\bibitem{Ebrahimi:2010:FVS:1736020.1736058}
E.~Ebrahimi, C.~J. Lee, O.~Mutlu, and Y.~N. Patt, ``{Fairness via Source
  Throttling: A Configurable and High-Performance Fairness Substrate for
  Multi-Core Memory Systems},'' in {\em Proceedings of the International
  Conference on Architectural Support for Programming Languages and Operating
  Systems (ASPLOS)}, pp.~335--346, 2010.

\bibitem{kim2010atlas}
Y.~Kim, D.~Han, O.~Mutlu, and M.~Harchol-Balter, ``{ATLAS: A Scalable and
  High-Performance Scheduling Algorithm for Multiple Memory Controllers},'' in
  {\em Proceedings of the International Symposium on High Performance Computer
  Architecture (HPCA)}, pp.~1--12, 2010.

\bibitem{tendler2002power4}
J.~M. Tendler, J.~S. Dodson, J.~Fields, H.~Le, and B.~Sinharoy, ``{POWER4
  System Microarchitecture},'' {\em IBM Journal of Research and Development},
  vol.~46, no.~1, pp.~5--25, 2002.

\bibitem{doweck2006inside}
J.~Doweck, ``{Inside Intel{\textregistered} Core Microarchitecture},'' in {\em
  IEEE Hot Chips Symposium (HCS)}, pp.~1--35, 2006.

\bibitem{Baer:1991:EOP:125826.125932}
J.-L. Baer and T.-F. Chen, ``{An Effective On-chip Preloading Scheme to Reduce
  Data Access Penalty},'' in {\em Proceedings of the ACM/IEEE Conference on
  Supercomputing}, pp.~176--186, 1991.

\bibitem{Sherwood:2000:PSB:360128.360135}
T.~Sherwood, S.~Sair, and B.~Calder, ``{Predictor-Directed Stream Buffers},''
  in {\em Proceedings of the International Symposium on Microarchitecture
  (MICRO)}, pp.~42--53, 2000.

\bibitem{Ishii:2009:AMP:1542275.1542349}
Y.~Ishii, M.~Inaba, and K.~Hiraki, ``{Access Map Pattern Matching for Data
  Cache Prefetch},'' in {\em Proceedings of the International Conference on
  Supercomputing (ICS)}, pp.~499--500, 2009.

\bibitem{Sair:2003:DPS:642791.642793}
S.~Sair, T.~Sherwood, and B.~Calder, ``{A Decoupled Predictor-Directed Stream
  Prefetching Architecture},'' {\em IEEE Transactions on Computers (TC)},
  vol.~52, pp.~260--276, March 2003.

\bibitem{Jouppi:1990:IDC:325164.325162}
N.~P. Jouppi, ``{Improving Direct-Mapped Cache Performance by the Addition of a
  Small Fully-Associative Cache and Prefetch Buffers},'' in {\em Proceedings of
  the International Symposium on Computer Architecture (ISCA)}, pp.~364--373,
  1990.

\bibitem{Palacharla:1994:ESB:191995.192014}
S.~Palacharla and R.~E. Kessler, ``{Evaluating Stream Buffers As a Secondary
  Cache Replacement},'' in {\em Proceedings of the International Symposium on
  Computer Architecture (ISCA)}, pp.~24--33, 1994.

\bibitem{Zhang:2000:HSP:335231.335247}
C.~Zhang and S.~A. McKee, ``{Hardware-Only Stream Prefetching and Dynamic
  Access Ordering},'' in {\em Proceedings of the International Conference on
  Supercomputing (ICS)}, pp.~167--175, 2000.

\bibitem{Iacobovici:2004:ESE:1006209.1006211}
S.~Iacobovici, L.~Spracklen, S.~Kadambi, Y.~Chou, and S.~G. Abraham,
  ``{Effective Stream-Based and Execution-Based Data Prefetching},'' in {\em
  Proceedings of the International Conference on Supercomputing (ICS)},
  pp.~1--11, 2004.

\bibitem{Somogyi:2006:SMS:1135775.1136508}
S.~Somogyi, T.~F. Wenisch, A.~Ailamaki, B.~Falsafi, and A.~Moshovos, ``{Spatial
  Memory Streaming},'' in {\em Proceedings of the International Symposium on
  Computer Architecture (ISCA)}, pp.~252--263, 2006.

\bibitem{Nesbit:2004:DCP:1072448.1072460}
K.~J. Nesbit and J.~E. Smith, ``{Data Cache Prefetching Using a Global History
  Buffer},'' in {\em Proceedings of the International Symposium on High
  Performance Computer Architecture (HPCA)}, pp.~96--, 2004.

\bibitem{Shevgoor:2015:EPC:2830772.2830793}
M.~Shevgoor, S.~Koladiya, R.~Balasubramonian, C.~Wilkerson, S.~H. Pugsley, and
  Z.~Chishti, ``{Efficiently Prefetching Complex Address Patterns},'' in {\em
  Proceedings of the International Symposium on Microarchitecture (MICRO)},
  pp.~141--152, 2015.

\bibitem{Nesbit:2004:AAD:1025127.1026003}
K.~J. Nesbit, A.~S. Dhodapkar, and J.~E. Smith, ``{AC/DC: An Adaptive Data
  Cache Prefetcher},'' in {\em Proceedings of the International Conference on
  Parallel Architectures and Compilation Techniques (PACT)}, pp.~135--145,
  2004.

\bibitem{Kumar:1998:ESL:279358.279404}
S.~Kumar and C.~Wilkerson, ``{Exploiting Spatial Locality in Data Caches Using
  Spatial Footprints},'' in {\em Proceedings of the International Symposium on
  Computer Architecture (ISCA)}, pp.~357--368, 1998.

\bibitem{Chen:2004:ACS:1072448.1072476}
C.~F. Chen, S.-H. Yang, B.~Falsafi, and A.~Moshovos, ``{Accurate and
  Complexity-Effective Spatial Pattern Prediction},'' in {\em Proceedings of
  the International Symposium on High Performance Computer Architecture
  (HPCA)}, pp.~276--287, 2004.

\bibitem{Cantin:2006:SP:1168857.1168892}
J.~F. Cantin, M.~H. Lipasti, and J.~E. Smith, ``{Stealth Prefetching},'' in
  {\em Proceedings of the International Conference on Architectural Support for
  Programming Languages and Operating Systems (ASPLOS)}, pp.~274--282, 2006.

\bibitem{bingo_dpc3}
M.~Bakhshalipour, M.~Shakerinava, P.~Lotfi-Kamran, and H.~Sarbazi-Azad,
  ``{Accurately and Maximally Prefetching Spatial Data Access Patterns with
  Bingo},'' {\em The Third Data Prefetching Championship}, 2019.

\bibitem{Joseph:1997:PUM:264107.264207}
D.~Joseph and D.~Grunwald, ``{Prefetching Using Markov Predictors},'' in {\em
  Proceedings of the International Symposium on Computer Architecture (ISCA)},
  pp.~252--263, 1997.

\bibitem{Chou:2007:LEC:1331699.1331727}
Y.~Chou, ``{Low-Cost Epoch-Based Correlation Prefetching for Commercial
  Applications},'' in {\em Proceedings of the International Symposium on
  Microarchitecture (MICRO)}, pp.~301--313, 2007.

\bibitem{wenisch2009practical}
T.~F. Wenisch, M.~Ferdman, A.~Ailamaki, B.~Falsafi, and A.~Moshovos,
  ``{Practical Off-Chip Meta-Data for Temporal Memory Streaming},'' in {\em
  Proceedings of the International Symposium on High Performance Computer
  Architecture (HPCA)}, pp.~79--90, 2009.

\bibitem{Wenisch:2005:TSS:1069807.1069989}
T.~F. Wenisch, S.~Somogyi, N.~Hardavellas, J.~Kim, A.~Ailamaki, and B.~Falsafi,
  ``{Temporal Streaming of Shared Memory},'' in {\em Proceedings of the
  International Symposium on Computer Architecture (ISCA)}, pp.~222--233, 2005.

\bibitem{Jain:2013:LIM:2540708.2540730}
A.~Jain and C.~Lin, ``{Linearizing Irregular Memory Accesses for Improved
  Correlated Prefetching},'' in {\em Proceedings of the International Symposium
  on Microarchitecture (MICRO)}, pp.~247--259, 2013.

\bibitem{bakhshalipour2017efficient}
M.~Bakhshalipour, P.~Lotfi-Kamran, and H.~Sarbazi-Azad, ``{An Efficient
  Temporal Data Prefetcher for L1 Caches},'' {\em IEEE Computer Architecture
  Letters (CAL)}, vol.~16, no.~2, pp.~99--102, 2017.

\bibitem{Somogyi:2009:SMS:1555754.1555766}
S.~Somogyi, T.~F. Wenisch, A.~Ailamaki, and B.~Falsafi, ``{Spatio-Temporal
  Memory Streaming},'' in {\em Proceedings of the International Symposium on
  Computer Architecture (ISCA)}, pp.~69--80, 2009.

\bibitem{Burcea:2008:PV:1346281.1346301}
I.~Burcea, S.~Somogyi, A.~Moshovos, and B.~Falsafi, ``{Predictor
  Virtualization},'' in {\em Proceedings of the International Conference on
  Architectural Support for Programming Languages and Operating Systems
  (ASPLOS)}, pp.~157--167, 2008.

\bibitem{dpc3}
``{The Third Data Prefetching Championship}.''
  \url{https://dpc3.compas.cs.stonybrook.edu/ }, 2019.

\bibitem{mlop_dpc3}
M.~Shakerinava, M.~Bakhshalipour, P.~Lotfi-Kamran, and H.~Sarbazi-Azad,
  ``{Multi-Lookahead Offset Prefetching},'' {\em The Third Data Prefetching
  Championship}, 2019.

\bibitem{sandbox}
S.~H. Pugsley, Z.~Chishti, C.~Wilkerson, P.-f. Chuang, R.~L. Scott, A.~Jaleel,
  S.-L. Lu, K.~Chow, and R.~Balasubramonian, ``{Sandbox Prefetching: Safe
  Run-Time Evaluation of Aggressive Prefetchers},'' in {\em Proceedings of the
  International Symposium on High Performance Computer Architecture (HPCA)},
  pp.~626--637, 2014.

\bibitem{bop}
P.~Michaud, ``{Best-Offset Hardware Prefetching},'' in {\em Proceedings of the
  International Symposium on High Performance Computer Architecture (HPCA)},
  pp.~469--480, 2016.

\bibitem{rmd}
F.~Golshan, M.~Bakhshalipour, M.~Shakerinava, A.~Ansari, P.~Lotfi-Kamran, and
  H.~Sarbazi-Azad, ``{Harnessing Pairwise-Correlating Data Prefetching with
  Runahead Metadata},'' {\em IEEE Computer Architecture Letters (CAL)}, 2020.

\bibitem{hu2002timekeeping}
Z.~Hu, S.~Kaxiras, and M.~Martonosi, ``{Timekeeping in the Memory System:
  Predicting and Optimizing Memory Behavior},'' in {\em Proceedings of the
  International Symposium on Computer Architecture (ISCA)}, pp.~209--220, 2002.

\bibitem{hu:tcp}
Z.~Hu, M.~Martonosi, and S.~Kaxiras, ``{TCP: Tag Correlating Prefetchers},'' in
  {\em Proceedings of the International Symposium on High Performance Computer
  Architecture (HPCA)}, pp.~317--326, 2003.

\bibitem{lai:deadblock}
A.-C. Lai, C.~Fide, and B.~Falsafi, ``{Dead-Block Prediction \& Dead-Block
  Correlating Prefetchers},'' in {\em Proceedings of the International
  Symposium on Computer Architecture (ISCA)}, pp.~144--154, 2001.

\bibitem{spracklen2005effective}
L.~Spracklen, Y.~Chou, and S.~G. Abraham, ``{Effective Instruction Prefetching
  in Chip Multiprocessors for Modern Commercial Applications},'' in {\em
  Proceedings of the International Symposium on High-Performance Computer
  Architecture (HPCA)}, pp.~225--236, 2005.

\end{thebibliography}

\end{document}